\documentclass{aastex63}
\usepackage{natbib}
\usepackage{threeparttable}
\usepackage{color}
\usepackage{enumerate}
\usepackage{graphicx}
\usepackage{epstopdf}
\usepackage{amsthm,amsmath,amssymb}
\usepackage{mathrsfs}

\usepackage{txfonts}

\shorttitle{Observations of SN 2016ije}
\shortauthors{Li, Z., Zhang, T., Wang, X., et al.}

\begin{document} 

\title{SN 2016ije: An SN 2002es-like Type Ia Supernova Exploded in a Metal-poor and Low-surface Brightness Galaxy}

\author[0000-0001-6489-163X]{Zhitong Li}
\affil{Key Laboratory of Optical Astronomy, National Astronomical Observatories, Chinese Academy of Sciences, Beijing 100101, China}
\affil{School of Astronomy and Space Science, University of Chinese Academy of Sciences, Beijing 101408, China}

\author[0000-0002-8531-5161]{Tianmeng Zhang}
\email{zhangtm@nao.cas.cn}
\affil{Institute for Frontiers in Astronomy and Astrophysics, Beijing Normal University, Beijing, 102206, China}
\affil{Key Laboratory of Space Astronomy and Technology, National Astronomical Observatories, Chinese Academy of Sciences, 20A Datun Road, Beijing 100101, China}
\affil{School of Astronomy and Space Science, University of Chinese Academy of Sciences, Beijing 101408, China}

\author[0000-0002-7334-2357]{Xiaofeng Wang}
\email{wang\_xf@mail.tsinghua.edu.cn}
\affil{Physics Department and Tsinghua Center for Astrophysics (THCA), Tsinghua University, Beijing, 100084, China}
\affil{Beijing Planetarium, Beijing Academy of Science and Technology, Beijing, 100044, China}

\author[0000-0002-8296-2590]{Jujia Zhang}
\affil{Yunnan Observatories, Chinese Academy of Sciences, Kunming 650216, China}
\affil{Key Laboratory for the Structure and Evolution of Celestial Objects, Chinese Academy of Sciences, Kunming 650216, China}

\author[0000-0002-1296-6887]{Lluís Galbany}
\affil{Institute of Space Sciences (ICE, CSIC), Campus UAB, Carrer de Can Magrans, s/n, E-08193 Barcelona, Spain}
\affil{Institut d’Estudis Espacials de Catalunya (IEEC), E-08034 Barcelona, Spain}
\author[0000-0003-3460-0103]{Alexei V. Filippenko}
\affil{Department of Astronomy, University of California, Berkeley, CA 94720-3411, USA}
\author[0000-0001-5955-2502]{Thomas G. Brink}
\affil{Department of Astronomy, University of California, Berkeley, CA 94720-3411, USA}
\author[0000-0002-5221-7557]{Chris Ashall}
\affil{Department of Physics, Virginia Tech, 850 West Campus Drive, Blacksburg VA, 24061, USA}
\author[0000-0002-2636-6508]{WeiKang Zheng}
\affil{Department of Astronomy, University of California, Berkeley, CA 94720-3411, USA}
\author[0000-0001-6069-1139]{Thomas de Jaeger}
\affil{Institute for Astronomy, University of Hawaii, 2680 Woodlawn Drive, Honolulu, HI 96822, USA}
\author{Fabio Ragosta}
\affil{INAF – Osservatorio Astronomico di Roma, via di Frascati 33, I-00078 Monte Porzio Catone (Roma), Italy}
\author{Maxime Deckers}
\affil{School of Physics, Trinity College Dublin, The University of Dublin, Dublin 2, Ireland}
\author[0000-0002-1650-1518]{Mariusz Gromadzki}
\affil{Astronomical Observatory, University of Warsaw, Al. Ujazdowskie 4, 00-478 Warszawa, Poland}
\author[0000-0002-1229-2499]{D. R. Young}
\affil{Astrophysics Research Centre, School of Mathematics and Physics, Queen’s University Belfast, Belfast BT7 1NN, UK}

\author{Gaobo Xi}
\affil{Physics Department and Tsinghua Center for Astrophysics (THCA), Tsinghua University, Beijing, 100084, China}
\author{Juncheng Chen}
\affil{Wu Zhou University, Wuzhou 543002, China}
\author[0000-0002-3204-2358]{Xulin Zhao}
\affil{Department of Physics, Tianjin University of Technology, Tianjin 300384, China}
\author{Hanna Sai}
\affil{Physics Department and Tsinghua Center for Astrophysics (THCA), Tsinghua University, Beijing, 100084, China}
\author{Shengyu Yan}
\affil{Physics Department and Tsinghua Center for Astrophysics (THCA), Tsinghua University, Beijing, 100084, China}
\author[0000-0002-1089-1519]{Danfeng Xiang}
\affil{Physics Department and Tsinghua Center for Astrophysics (THCA), Tsinghua University, Beijing, 100084, China}
\author{Zhihao Chen}
\affil{Physics Department and Tsinghua Center for Astrophysics (THCA), Tsinghua University, Beijing, 100084, China}
\author{Wenxiong Li}
\affil{The School of Physics and Astronomy, Tel Aviv University, Tel Aviv 69978, Israel}
\author[0000-0002-3231-1167]{Bo Wang}
\affil{Yunnan Observatories, Chinese Academy of Sciences, Kunming 650216, China}
\affil{Key Laboratory for the Structure and Evolution of Celestial Objects, Chinese Academy of Sciences, Kunming 650216, China}

\author[0000-0002-6684-3997]{Hu Zou}
\affil{Key Laboratory of Optical Astronomy, National Astronomical Observatories, Chinese Academy of Sciences, Beijing 100101, China}
\author{Jipeng Sui}
\affil{Key Laboratory of Optical Astronomy, National Astronomical Observatories, Chinese Academy of Sciences, Beijing 100101, China}
\affil{School of Astronomy and Space Science, University of Chinese Academy of Sciences, Beijing 101408, China}
\author{Jiali Wang}
\affil{Key Laboratory of Optical Astronomy, National Astronomical Observatories, Chinese Academy of Sciences, Beijing 100101, China}
\author[0000-0001-6329-6644]{Jun Ma}
\affil{Key Laboratory of Optical Astronomy, National Astronomical Observatories, Chinese Academy of Sciences, Beijing 100101, China}
\affil{School of Astronomy and Space Science, University of Chinese Academy of Sciences, Beijing 101408, China}
\author[0000-0001-6590-8122]{Jundan Nie}
\affil{Key Laboratory of Optical Astronomy, National Astronomical Observatories, Chinese Academy of Sciences, Beijing 100101, China}
\author[0000-0002-0660-0432]{Suijian Xue}
\affil{Key Laboratory of Optical Astronomy, National Astronomical Observatories, Chinese Academy of Sciences, Beijing 100101, China}
\author{Xu Zhou}
\affil{Key Laboratory of Optical Astronomy, National Astronomical Observatories, Chinese Academy of Sciences, Beijing 100101, China}
\author{Zhimin Zhou}
\affil{Key Laboratory of Optical Astronomy, National Astronomical Observatories, Chinese Academy of Sciences, Beijing 100101, China}

\begin{abstract}

We have conducted photometric and spectroscopic observations of the peculiar Type Ia supernova (SN Ia) 2016ije that was discovered through the Tsinghua-NAOC Transient Survey. This peculiar object exploded in the outskirts of a metal-poor, low-surface brightness galaxy (i.e., $M_{g}$ = $-$14.5 mag). Our photometric analysis reveals that SN~2016ije is subluminous ($M_{B,\rm{max}}$ = $-$17.65$\pm$0.06 mag) but exhibits relatively broad light curves (${\Delta}m_{15}(B)$ = 1.35$\pm$0.14 mag), similar to the behavior of SN~2002es. Our analysis of the bolometric light curve indicates that only 0.14$\pm$0.04 $M_{\odot}$ of $^{56}$Ni was synthesized in the explosion of SN~2016ije, which suggests a less energetic thermonuclear explosion when compared to normal SNe~Ia, and this left a considerable amount of unburned materials in the ejecta. 
Spectroscopically, SN~2016ije resembles other SN~2002es-like SNe~Ia, except that the ejecta velocity inferred from its carbon absorption line ($\sim$ 4500~km~s$^{-1}$) is much lower than that from silicon lines ($\sim$ 8300~km~s$^{-1}$) at around the maximum light. Additionally, most of the absorption lines are broader than other 02es-like SNe Ia. These peculiarities suggest the presence of significant unburned carbon in the inner region and a wide line-forming region along the line of sight. These characteristics suggest that SN 2016ije might originate from the violent merger of a white dwarf binary system, when viewed near an orientation along the iron-group-element cavity caused by the companion star.
\end{abstract}


\section{Introduction}\label{sec:intro}

It is generally accepted that Type Ia supernovae (SNe~Ia) arise from thermonuclear explosions of carbon-oxygen white dwarfs (C-O WDs) in close binary systems. Although the question of whether the companions in the progenitor systems are degenerate WDs or nondegenerate stars remains controversial \citep{wb12, wxf13b, maoz14, wb18}, and the explosion models are inconclusive, the relatively uniform observational properties of SNe~Ia have led to their widespread use in cosmological studies. Some of the empirical relationships that have been found between their peak luminosities and the width of light/color curves (e.g., \citealt{phi93,guy05,wxf05,burns14}) make them good distance indicators. Large samples of SNe~Ia have been used to estimate the Hubble constant (e.g., \citealt{ham96a,sand06,riess21}) and to determine the expansion history of the universe \citep{riess98,per99,bet14}. 

Relative homogeneity in photometric and spectral evolution is common in SNe~Ia ($\sim$70\%; \citealt{liw11a}). These SNe~Ia have been known as ``Branch-normal'' ones \citep{bran93, fili97}. In contrast to the Branch-normal SNe~Ia, some SNe~Ia are classified into different subclasses because they show different photometric or spectral evolution. The overluminous group, such as SN~1991T, are characterized by broad and luminous light curves, relatively weak Si~{\sc ii} $\lambda$6355, and prominent Fe~{\sc ii/iii} absorption features around maximum light \citep{fili92a, phi92, ruiz92}. In contrast, subluminous SNe~Ia, such as SN~1991bg (hereafter ``91bg-like'') showed fast-evolving light curves and prominent absorption lines of intermediate-mass elements (IMEs; \citealt{fili92b,lei93}). Another peculiar subclass of SNe~Ia is the SN~2002es-like events (hereafter ``02es-like''), which are similar to 91bg-like SNe~Ia in terms of low luminosity and spectral characteristics but display broader light curves \citep{02es,whi15}. 

Although the first 02es-like SN was observed at the end of the 1990s \citep{ald99}, this group was defined and established more recently \citep{02es,whi15}. In general, it refers to SNe~Ia similar to SN~2002es, which has $BV$-band light curves comparable to those of normal SNe~Ia but with peak magnitudes similar to those of 91bg-like SNe~Ia. The optical spectra of SN~2002es also have similar properties to those of the 91bg-like subclass, characterized by strong Si~{\sc ii} $\lambda$5972, O~{\sc i}, and Ti~{\sc ii} lines near maximum light. Strictly speaking, 02es-like SNe~Ia should have low ejecta velocities ($\sim 6000$~km~s$^{-1}$) similar to those of SN~2002es, but this is not required in a wider definition \citep{tau17}. For example, SN~2006bt and PTF10ops with near-maximum-light Si~{\sc ii} velocities ($v_{\rm Si}$) $\approx$ 10,000~km~s$^{-1}$ and iPTF14atg with $v_{\rm Si} \approx 8000$~km~s$^{-1}$ are also included in the 02es-like subclass.

In addition to these commonalities and diversity in ejecta velocities, observations of 02es-like SNe~Ia have revealed other properties. An ultraviolet (UV) spike was detected in the early-time observations of the 02es-like SN~Ia iPTF14atg \citep{cao16}. SN~2019yvq with an early UV excess was also classified as a transitional member of the 02es-like subclass \citep{bur21}, but SN~2019yvq is more similar to a high-velocity object given its spectral features \citep{wxf09a}. It is not clear whether the UV spike is common for this peculiar group owing to the insufficient sample. The 02es-like object SN~2010lp did not display [Fe~{\sc iii}] but rather low-velocity [O~{\sc i}] \citep{tau13} in its nebular spectrum. Owing to the limited sample with nebular-phase spectra, it is still unknown whether this behavior is common for 02es-like SNe~Ia. Note that the light curves of SN~2002es exhibited a rapid decline from $t \approx 1$ month after the peak \citep{02es}, which is unique among 02es-like SNe~Ia. Among the current sample of 02es-like SNe~Ia, SN~2006bt and PTF10ops are located far from their host galaxies and some tend to occur in red galaxies \citep{whi15}, which suggests that they likely arise from old stellar populations.

To explain the photometric properties of 02es-like SNe~Ia, \citet{mag11} and \citet{02es} suggested that a violent merger model of two 0.9 $M_{\odot}$ WDs \citep{pak10} would be a promising scenario. This model was originally proposed to explain the 91bg-like SNe~Ia, but the predicted light curves are too broad and they are more consistent with the photometric properties of 02es-like SNe~Ia. \citet{kro13,kro16} updated the model with a merger of 0.9 $M_{\odot}$ and 0.76 $M_{\odot}$ WDs at different metallicities, from which the produced spectra and light curves are in good agreement with those of 02es-like SNe Ia such as SN 2010lp and iPTF14atg.  

Nevertheless, interpretations of the observed properties of 02es-like SNe~Ia with the violent merger model remain controversial. \citet{kro13} suggested that the low central density of the violent merger model can explain the deficiency of [Fe~{\sc iii}], as well as the abundance of [O~{\sc i}] in the nebular-phase spectra of 02es-like SNe~Ia. However, \citet{maz22} pointed out that the [O~{\sc i}] line of SN~2010lp has an unusual double peak, which is hard to reproduce by a violent merger model. Although \citet{cao16} ruled out many explanations for the early-time UV spike of iPTF14atg except for the ejecta-companion interaction, \citet{kro16} argued that interaction with compact or aspherical circumstellar matter (CSM) could not be excluded. Moreover, \citet{lev17} fit the early light excess of iPTF14atg with the disc-originated matter (DOM) model, which suggests that the progenitor of iPTF14atg could be a double-degenerate system.

In this paper, we present photometric and spectroscopic observations of SN~2016ije, which is a member of the subclass of 02es-like SNe~Ia. Observations, data reduction, and estimates of the properties of the host galaxy are described in Section \ref{sec:observation}. Section \ref{sec:photometric} presents the light and color curves, as well as the bolometric light curve. In Section \ref{sec:spectra}, we describe the spectroscopic evolution and quantify the properties of important spectral lines. The peculiarities of the spectra, together with comparisons with different explosion models, are discussed in Section \ref{sec:discussion}. We summarize this work in Section \ref{sec:conclusion}.

\section{Observations and Data Reduction}\label{sec:observation}

\begin{figure}[ht]
\includegraphics[angle=0,width=160mm]{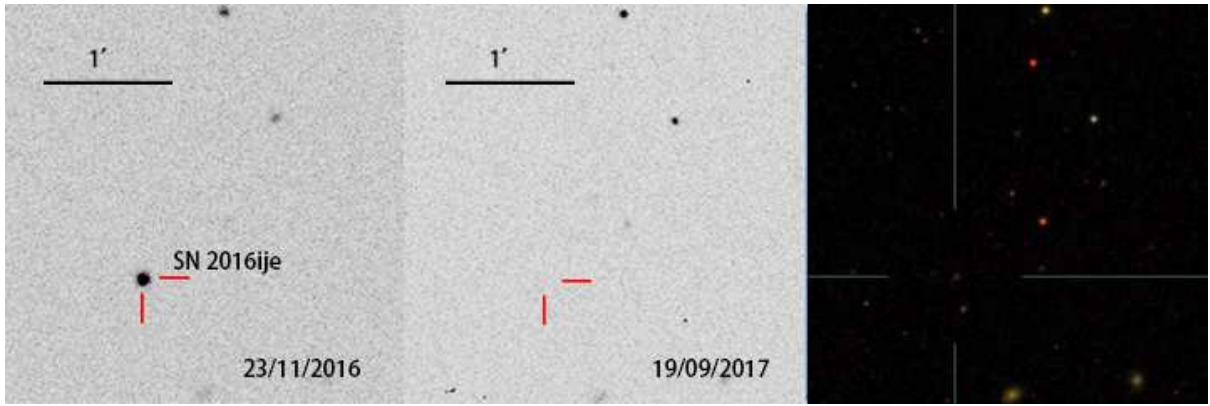}
\centering
\caption {Left-hand panel: the unfiltered image of SN~2016ije taken by TNTS on 2016 November 23. Middle panel: the unfiltered image of the same area taken by TNTS on 2017 Sep. 09, when SN~2016ije had faded away. Right-hand panel: deeper image taken by SDSS, in which we can find the faint host galaxy. The position of the SN is marked by the cross-hair.}
\label{fig:findingchart}
\end{figure}

SN~2016ije was discovered on 2016 November 22.71 (UT dates are used throughout this paper) by the 0.6~m Schmidt telescope in the course of the THU-NAOC\footnote{National Astronomical Observatories, Chinese Academy of Sciences} Transient Survey (TNTS; \citealt{zhangtm15}). The unfiltered magnitude of the discovery is reported as 17.8 mag. The coordinates are $\alpha$ = 01$^{\rm{h}}$58$^{\rm{m}}$30$^{\rm{s}}$.33, $\delta$ = +12$^{\circ}$55$\arcmin$27$\arcsec$.9 (J2000), located 1$\arcsec$.0 east of the center of the faint, tiny galaxy SDSS J015830.25+125528.1. No redshift was reported for this galaxy. Figure \ref{fig:findingchart} shows the finder chart of SN~2016ije. One day after the discovery, an optical spectrum was taken with the Lijiang 2.4~m telescope (LJT; \citealt{LJT,YFOSC}) of Yunnan Observatories (YNAO), which was used to classify SN~2016ije as a peculiar SN~Ia \citep{zhangjj16}.

\subsection{Photometry}

After the discovery, we performed follow-up optical and near-infrared (NIR) observations of SN~2016ije on several facilities. The optical photometry ($BVRI$) was obtained mainly by the 0.8~m Tsinghua-NAOC Telescope (TNT\footnote{This telescope is operated by Tsinghua University and NAOC.}) that is located at Xinglong Observatory of NAOC \citep{wxf08}. It is equipped with a 1340 $\times$ 1300 pixel back-illuminated CCD, with a field of view (FoV) of 11.5$\arcmin$ $\times$ 11.2$\arcmin$ (pixel size $\sim$ 0.52$\arcsec$ pixel$^{-1}$; \citealt{huang12}). The TNT instrumental magnitudes were obtained using an {\it ad hoc} pipeline (based on the IRAF\footnote{IRAF, the Image Reduction and Analysis Facility, is distributed by the National Optical Astronomy Observatories, which are operated by the Association of Universities for Research in Astronomy (AURA), Inc., under cooperative agreement with the National Science Foundation.} DAOPHOT package; \citealt{stet87}).
Since SN~2016ije exploded near the center of the host galaxy, we applied image subtraction with the template images taken $\sim 300$ days after the explosion when the SN had faded away. 

Some $UBVRI$-band photometry was obtained by the Lijiang 2.4~m Telescope (LJT; \citealt{LJT,YFOSC}) and was reduced using the standard point-spread function (PSF) fitting method from the IRAF DAOPHOT package \citep{stet87}. Note the template image for SN~2016ije was not taken by LJT, thus the magnitudes may be overestimated,  especially at late time, but the host-galaxy contamination is limited \footnote{\text We estimated that the t $\sim$ 50 day photometry by LJT suffered a host-galaxy contamination of less than 0.2 mag in the $V$ band.}. Moreover, the PSF photometry can help alleviate the host-galaxy contamination. Some $BVRIJHK$-band photometry was also obtained through the PESSTO project \citep{sma15}, with the EFOSC \citep{EFOSC} and SOFI \citep{SOFI} mounted on the New Technology Telescope (NTT) at La Silla Observatory, and was reduced using the PESSTO data-reduction pipeline \citep{sma15}. 

The optical magnitudes were calibrated using standard stars and adjusted with zero-point correction. The $S$- or $K$- corrections were not applied to the photometry. The magnitudes of the standard stars were obtained from \textbf{Sloan} catalogs and converted into those of the Landolt system by applying the equations provided by \citet{jes05}. The final calibrated optical magnitudes of SN~2016ije are presented in Table \ref{table:photometry}. The $JHK$-band magnitudes calibrated by the 2MASS catalog are presented in Table \ref{table:photometry_nir}. The photometric error includes the photon noise of SN~2016ije and the statistic error of calibration by standard stars.

\subsection{Spectroscopy}

Eleven spectra of SN~2016ije were collected with the NTT, LJT, and Xinglong 2.16~m Telescope (XLT; \citealt{XLT216}), including 10 optical spectra and one NIR spectrum. The spectroscopic observations covered phases from $t \approx -0.9$ to +54.4 days with respect to peak brightness. In addition, one optical spectrum was obtained with the Kast spectrograph \citep{miller-stone93} on the 3~m Shane reflector at Lick Observatory at $t \approx 6.2$ days. The $2\arcsec$-wide slit was aligned near the parallactic angle \citep{fil82} to minimize differential light losses caused by atmospheric dispersion. The spectra of NTT and LJT were also taken with the silts aligned along the parallactic angles. While this is not the case in the XLT observations, the spectra of SN and standard stars were obtained under similar conditions (i.e., with the same positions on the slit and similar altitude and time), which can help to mitigate the impact of differential flux loss. A journal of spectroscopic observations is given in Table \ref{table:spec}.

We reduced all of the optical spectra taken by LJT, XLT, and Shane using standard IRAF routines, including corrections for bias, flat field, and removal of cosmic rays. The wavelength scale of the spectra was calibrated by arc-lamp spectra and the flux was calibrated using standard stars that were observed on the same night at a similar airmass as the SN. All of the spectra were corrected for atmospheric extinction using mean extinction curves that were obtained at each observatory, and telluric lines were also removed whenever possible. Spectra taken by the NTT were fully reduced and calibrated using the PESSTO data-reduction pipeline \citep{sma15}. Because of its poor signal-to-noise ratio, the NIR spectrum is not used in subsequent analysis.

\subsection{Host Galaxy} \label{sec:host}

\begin{figure}[ht]
\centering
\includegraphics[angle=0,width=120mm]{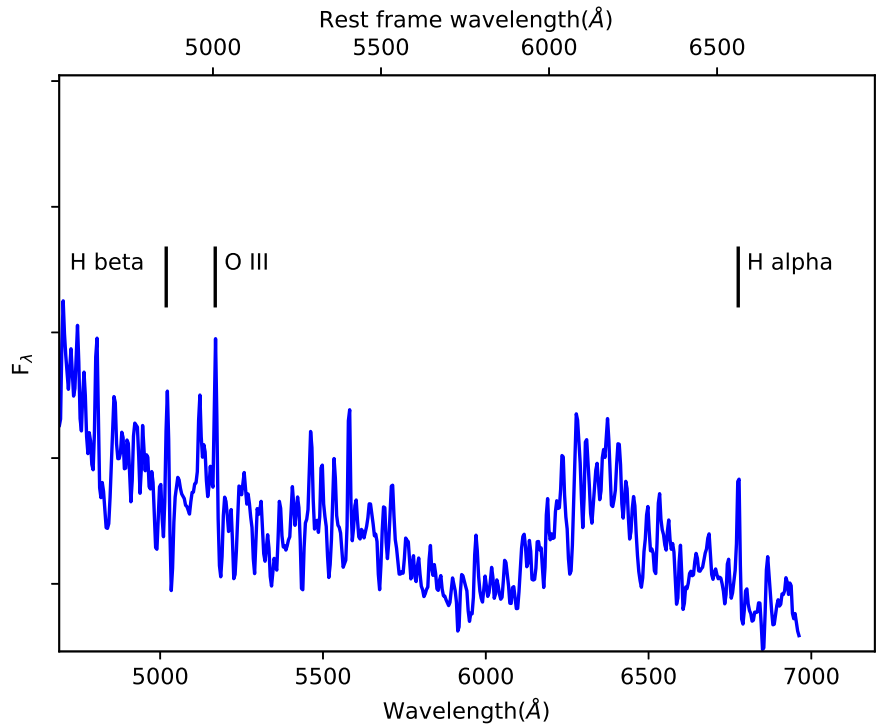}
\caption{The spectrum of the host galaxy SDSS J015830.25+125528.1 taken by Keck II + DEIMOS. Parts of the continuum shape are not reliable (e.g., the broad bump near 6300~\AA). The H$\beta$, [O~{\sc iii}], and H$\alpha$ emission lines at 4864.35, 5169.92, and 6774.43~{\AA} are marked, from which the redshift was determined to be $z$ = 0.0324 $\pm$ 0.0003.}
\label{fig:host}
\end{figure}

To determine the redshift of the faint host galaxy of SN~2016ije, we used Keck~II + DEIMOS \citep{fab03} to take a spectrum of SDSS J015830.25+125528.1 on 2017 October 14.59. The host-galaxy spectrum is shown in Figure \ref{fig:host}, where two prominent emission lines are visible at 5169.92 and 6774.43~{\AA}. Assuming that they are [O~{\sc iii}]~$\lambda$5007 and H$\alpha$, respectively, then we get a consistent redshift of $z$ = 0.0324 $\pm$ 0.0003. The corresponding luminosity distance is $d_L$ = 136.5 $\pm$ 2.9 Mpc and the distance modulus is $m - M$ = 35.68 $\pm$ 0.04 mag when adopting cosmological parameters $\Omega_{m} = 0.27$, $\Omega_\Lambda = 0.73$, and H$_0$ = 73 $\pm$ 1.4 km s$^{-1}$ Mpc$^{-1}$ \citep{riess21}.

With this distance, and the $grz$-band images and catalogs from the Sloan Digital Sky Survey (SDSS; \citealt{SDSS}), we derive the absolute magnitude of SDSS J015830.25+125528.1 to be $-$14.5 mag in $g$, which is much fainter than most host galaxies of 02es-like SNe ($-$19 to $-$22 mag; \citealt{whi15}). The $g$-band surface brightness is estimated to be 24.1 mag arcsec$^{-2}$ (\citealt{du15}, Eq. 1a), which indicates that the host galaxy has low surface brightness. The color $g-i = 0.76$ mag is bluer than most of the host galaxies of 02es-like SNe ($g-i = 1.3$ mag; \citealt{whi15}). 

To further quantify the properties of the host galaxy, we measure the intensity ratio of flux between [N~{\sc ii}]~$\lambda$6583 and H$\alpha$ as log([N~{\sc ii}]/H$\alpha$) $= -0.70$ and the ratio between [O~{\sc iii}]~$\lambda$5007 and H${\beta}$ as log([O~{\sc iii}]/H$\beta$) = 0.031, consistent with those of star-forming regions in the diagrams of \citet{bal81} and \citet{vei87}. Moreover, with the flux measurements of these four emission lines, the metallicity of the host galaxy was estimated to be 12 $+$ log(O/H) $=$ 8.46 $\pm$ 0.16 using an empirical relationship (\citealt{kew08}, Eq. A9). By fitting the spectrum with combinations of single-burst stellar population models using Firefly \citep{wil17}, we obtained a stellar mass of $M_{*} \approx 2.0 \times 10^{7}~M_{\odot}$, an age of 1.54~Gyr, and metallicity of the stellar populations of [Z/H] $= -0.37$. The low mass of the host galaxy can explain its low luminosity. The metallicities that were determined from different approaches are roughly consistent, which indicates that the host galaxy has half the solar metallicity. Note that the star-formation rate (SFR) can be estimated to be $2.4 \times 10^{-3}~M_{\odot}$~yr$^{-1}$ (\citealt{ken98}, Eq. 2). The stellar mass and the SFR of the host galaxy agree with a star-forming galaxy, and the specific star-formation rate (SSFR) is $1.2 \times 10^{-10}$~yr$^{-1}$.

Combining the spectral properties with the color and luminosity derived from SDSS, we consider the host galaxy of SN~2016ije to be a low-mass, subsolar-metallicity, star-forming galaxy. Since neither the spectra of SN~2016ije nor that of the host galaxy show any visible absorption feature of Na~{\sc i}~D, the extinction of the host galaxy is assumed to be negligible. The Galactic reddening toward SN~2016ije is taken to be $A_{V}^{\rm Gal}$ = 0.143~mag \citep{schla11}. 

\section{Photometric Properties} \label{sec:photometric}

\subsection{Optical Light Curves}\label{sec:opticalLC}

\begin{figure}[ht]
\centering
\includegraphics[angle=0,width=160mm]{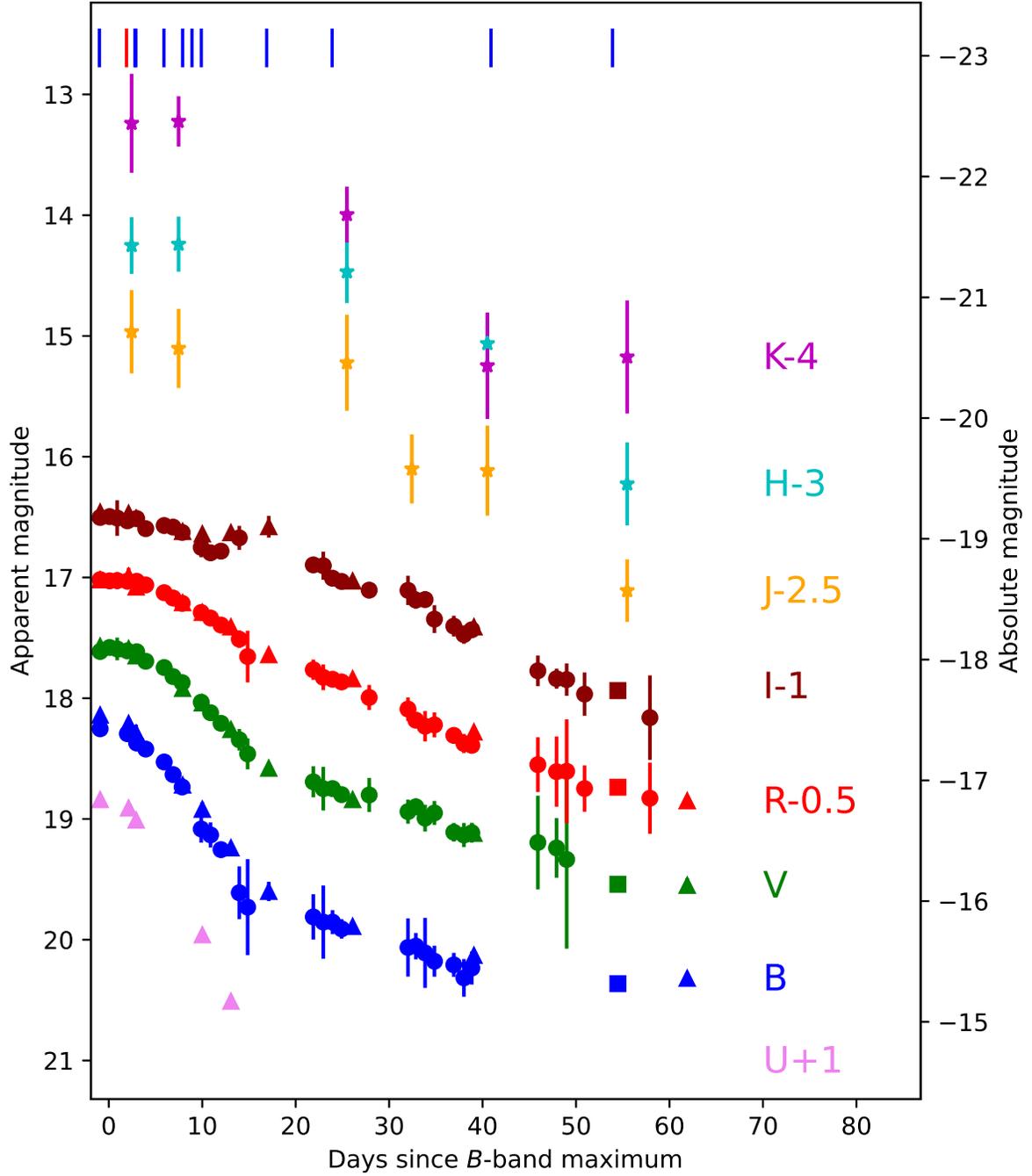}
\caption{$UBVRIJHK$ light curves of SN~2016ije obtained by TNT (circles), LJT (triangles), and NTT (squares and stars). The lines on the top of the figure indicate the phases when optical (blue) or near-infrared (red) spectra were taken.}
\label{fig:LC16ije}
\end{figure}

Figure \ref{fig:LC16ije} shows the $UBVRI$ light curves of SN~2016ije. Fitting them to those of SN~2002es near maximum light indicates that SN~2016ije reached $B$-band maximum on MJD 57716.6 $\pm$ 0.8 at $m_{B,\rm{max}}$ = 18.22 $\pm$ 0.04 mag. With these values, we estimate the post-peak decline rate to be ${\Delta}m_{15}(B)$ = 1.35 $\pm$ 0.14 mag. Adopting the reddening and distance modulus derived in Section \ref{sec:host}, the peak absolute magnitude of SN~2016ije is $M_{B,\rm{max}}$ = $-$17.65 $\pm$ 0.06 mag, which is similar to other 02es-like SNe~Ia. The relatively large uncertainty in the estimated peak time is due to insufficient observations before $B$-band maximum. The main photometric parameters of SN~2016ije are listed in Table \ref{table:obj}. For subsequent discussions, the phases are referred with respect to $B$-band maximum. 

\begin{figure}[ht]
\centering
\includegraphics[angle=0,width=160mm]{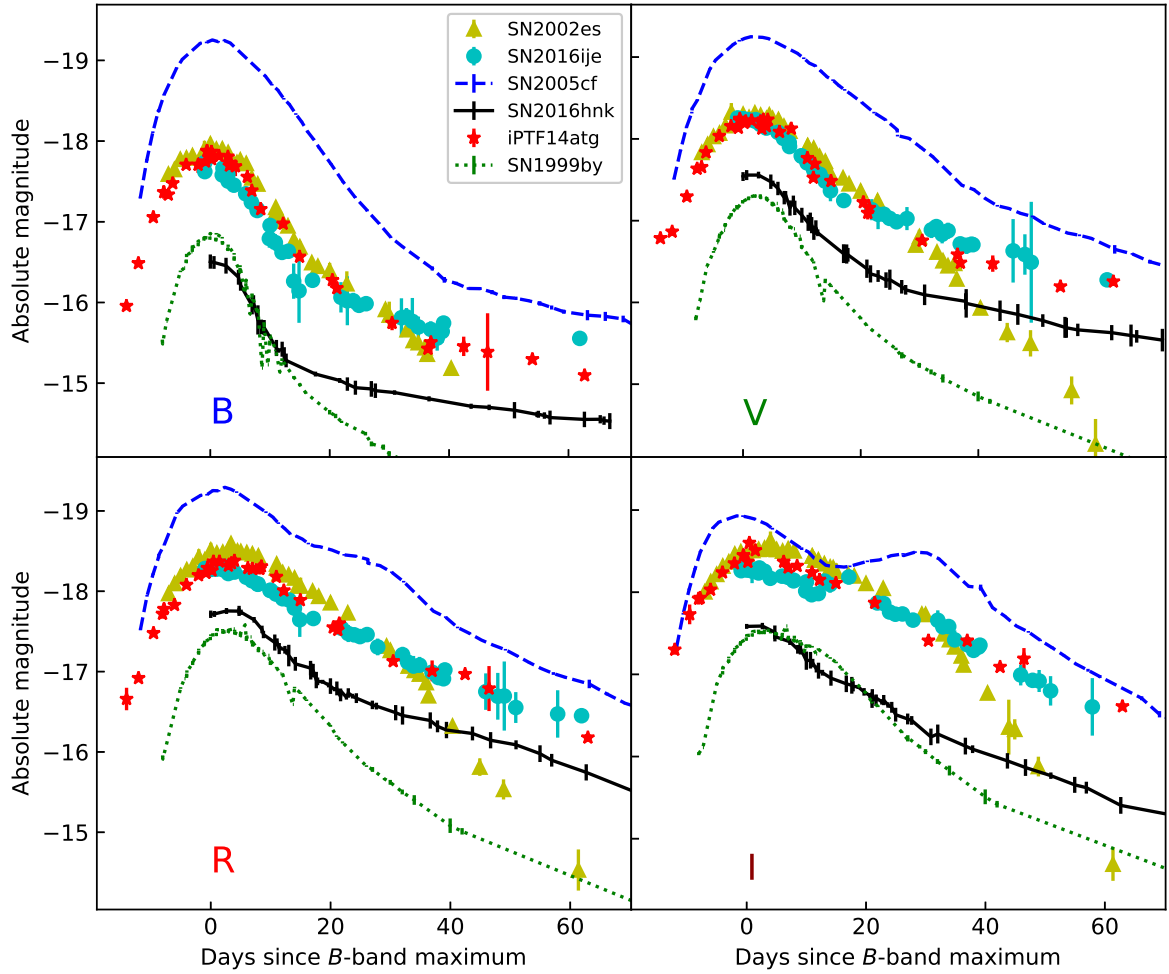}
\caption{Comparison of $BVRI$ light curves of SN~2016ije with those of SNe 1999by, 2002es, 2005cf, iPTF14atg, and 2016hnk. All of the extinctions were corrected. The $RI$-band light curves of iPTF14atg and SN~2016hnk are converted from their $ri$-band light curves with zero-point transformation for comparison purposes.}
\label{fig:lc_compare}
\end{figure}

In Figure \ref{fig:lc_compare}, we compare the $BVRI$ light curves of SN~2016ije with those of other normal and subluminous SNe~Ia, including SNe 1999by \citep{gar04,gane10}, 2002es \citep{02es}, 2005cf \citep{past07,wxf09b}, iPTF14atg \citep{cao16}, and 2016hnk \citep{gal19}. Of them, SN~1999by is a 91bg-like SN~Ia and SN~2005cf is a normal SN~Ia. SNe~2002es and iPTF14atg can be put into the subclass of 02es-like SNe~Ia, while SN~2016hnk is a peculiar subluminous SN~Ia with strong calcium features. All of the light curves have been corrected for Galactic and host-galaxy extinctions whenever possible. Since iPTF14atg and SN~2016hnk do not have $R$- and $I$-band observations, their $RI$-band light curves are converted from their $ri$-band light curves with zero-point transformation.

In terms of absolute peak magnitudes and post-peak declines, SN~2016ije is quite similar to SN~2002es and iPTF14atg. However, the $B$ and $V$ light curves of SN~2016ije decline slightly more slowly than those of iPTF14atg after $t \approx +20$ days, which is similar to the behavior seen in SN~2016hnk. In comparison with SN~2016ije and iPTF14atg, the post-peak light curves of SN~2016hnk start flattening at an even earlier phase, $t \approx 10$ days after the peak, which might be due to light scattering caused by dense dust that evolved from its companion star \citep{gal19}. However, SN~2016ije has no apparent Na~{\sc i}~D absorption feature in its spectra; thus, dense dust almost certainly does not exist around SN~2016ije, as suggested for SN~2016hnk.

The peak luminosities of the $RI$-band light curves of SN~2016ije are also similar to those of SN~2002es. Moreover, SN~2016ije and other subluminous SNe~Ia do not exhibit a secondary shoulder or bump in $R$ as do normal SNe~Ia. The weak $I$-band secondary peak appears at $t \approx 18$ days after the peak, which revealed the approximate time of iron-group elements (IGEs) recombination \citep{kas06,burns14,wyg19}. The secondary peak is not prominent in the $I$-band light curve of SN 2002es.

\subsection{Color Curves}\label{sec:color}

\begin{figure}[ht]
\centering
\includegraphics[angle=0,width=100mm]{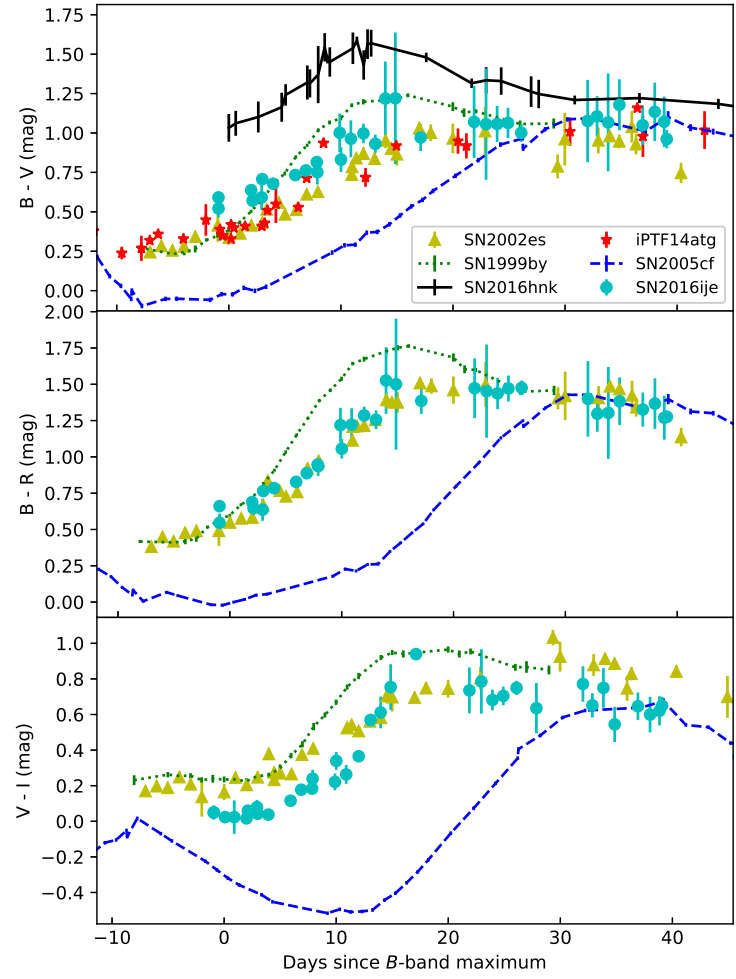}
\caption{Multi-band color curves of SN~2016ije, compared to those of SNe 1999by, 2002es, 2005cf, iPTF14atg, and 2016hnk. All of the extinctions were corrected.}
\label{fig:color}
\end{figure}

The reddening-corrected color curves of SN~2016ije, together with those of some subluminous and normal SNe~Ia for comparison, are shown in Figure \ref{fig:color}. Overall, the color-curve evolution of SN~2016ije is similar to those of SN~2002es and iPTF14atg. We noticed that the $B-V$ color curve of SN~2016hnk appears to be much redder than those of 02es-like SNe~Ia, such as SN~2016ije, even when accounting for a significant amount of reddening in the former case \citep{gal19}.

Using the color curve of SN 2002es as a template, we find that the $B-V$ color curve of SN~2016ije reached the reddest value at $t=+17.1\pm1.9$ days (i.e., the color stretch factor $s_{\rm{BV}}$ = 0.57 $\pm$ 0.06; \citealt{burns14}), which suggests that SN 2016ije reached the reddest color at about one day later than other 02es-like SNe~Ia. This phenomenon is consistent with the slower post-peak decline of the light curves, which suggests that the photosphere of SN~2016ije evolves slightly slower than that of other 02es-like SNe~Ia.

Moreover, we also notice that the color curves of 02es-like SNe~Ia follow a distinct evolution pattern relative to those of 91bg-like or normal SNe~Ia. The color curves of 02es-like SNe evolve similarly to those of other SNe~Ia before $t \approx +20$ days, though the former ones turn red slower and are redder than normal SNe Ia throughout this phase. After $t \approx +30$ days, the $B-V$ color evolution of 91bg-like and normal SNe~Ia is roughly in accordance with that predicted by the Lira~-~Phillips relation (i.e., with a slope of $-$0.0118 mag day$^{-1}$; \citealt{phi99} during t $\sim$ 30 - 90 days after maximum light). In contrast, the color of 02es-like SNe~Ia stays red until at least $t \approx +40$ days. This trend also exists in the color curve of the controversial 02es-like object SN 2006bt. The difference in color evolution can help us to distinguish borderline members of the 02es-like subclass, such as SN 2006bt, from other SNe~Ia. It also indicates the peculiarity in the photospheric evolution of 02es-like SNe.

\subsection{Bolometric Light Curves, and NIR and UV Photometry }\label{sec:NIRLC}\label{sec:bolo}

\begin{figure}[ht]
\centering
\includegraphics[angle=0,width=160mm]{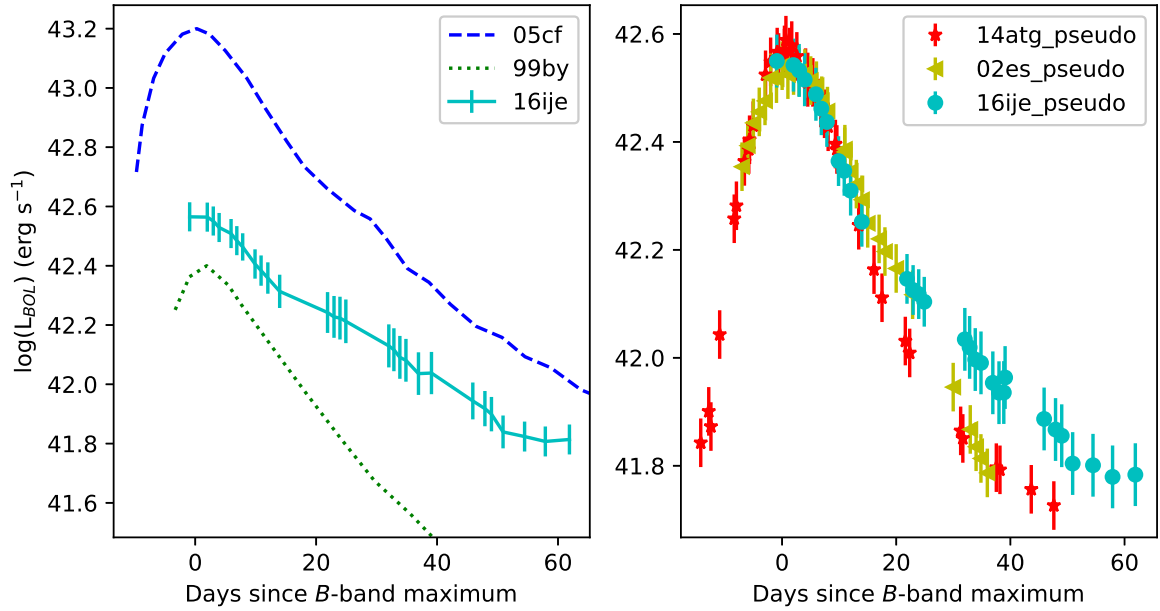}
\caption{Left-hand panel: bolometric ($UBVRIJHK$) light curve of SN~2016ije compared with those of SNe 1999by and 2005cf. Right-hand panel: pseudo-bolometric ($UBVRI$) light curves of SNe~2002es, 2016ije, and iPTF14atg, which do not contain the near-infrared flux, are also plotted for comparison.}
\label{fig:bol}
\end{figure}

We construct the bolometric light curve of SN~2016ije via trapezoidal integration of flux in the $UBVRIJHK$ bands at a wavelength from 3660 to 21,900 {\AA} (i.e., the equivalent wavelength of $U$ and $K$ bands). We estimate the UV missing flux of the bolometric luminosity of SN 2016ije (i.e., $\lambda$ \textless 3660 {\AA}) near the maximum light as 1.6\% of the total flux by the spectral energy distribution (SED) of iPTF14atg. Therefore, the contribution of the UV flux to the bolometric luminosity of SN 2016ije can be ignored near its maximum light. Due to the absence of late-time $U$-band observation, we adopt an extrapolation while assuming that the $U-B$ color remains constant to estimate the magnitude in the $U$ band. A comparison of bolometric light curves between SNe~2016ije, 1999by, and 2005cf is shown in Figure \ref{fig:bol}.

The pseudo-bolometric light curve of iPTF14atg does not include the NIR observations \citep{kro16}, so we also plot the pseudo-bolometric light curve of SNe 2002es and 2016ije of similar wavelengths for comparison in Figure \ref{fig:bol}. To correct the pseudo-bolometric light curve of SN 2002es without $U$-band observations, we adopt the $U$-band correction applied for SN 2016ije. However, since the pre-maximum $U$-band observations are unavailable for SN 2016ije, we have assumed that the proportion of $U$-band emission is constant during this phase. Although this assumption could potentially affect the bolometric light curve, we believe that such an impact is small because the $U$-band proportion is only about 10\%. The pseudo-bolometric light curve of SN 2016ije evolves more slowly than those of the other two 02es-like SNe Ia, although their peak luminosities are almost the same.
One can notice that the bolometric light curve of SN~2016ije shows a bump at $t \approx +30$ days when compared with those of SN~2002es and iPTF14atg. Note that the bump feature in SN 2016ije exhibits significant uncertainties and the bump is not present in its pseudo-bolometric light curve. This discrepancy is likely to be due to a few $JHK$ observations with large uncertainties at similar epochs.

From the bolometric light curve, we derive the peak luminosity of SN~2016ije as $L_{\rm{bol}}$ = (3.67 $\pm$ 0.47) $\times$ 10$^{42}$ erg s$^{-1}$. According to the radioactive decay law that was put forward by \citet{arn82}, we could estimate the mass of radioactive $^{56}$Ni synthesized in the explosion. However, the rise time of SN~2016ije cannot be tightly constrained because of the lack of pre-maximum observations. Since SN~2016ije is similar to SN~2002es in light-curve evolution, we estimate the same rise time as the latter, 16 $\pm$ 3 days \citep{02es}. With the $\gamma$ factor being adopted as 1.2 $\pm$ 0.2 \citep{nug95}, the nickel mass of SN~2016ije is estimated to be $M(^{56}\rm{Ni})$ = 0.14 $\pm$ 0.04 $M_{\odot}$.

\section{Spectral Properties}\label{sec:spectra}

\begin{figure}[ht]
\centering
\includegraphics[angle=0,width=140mm]{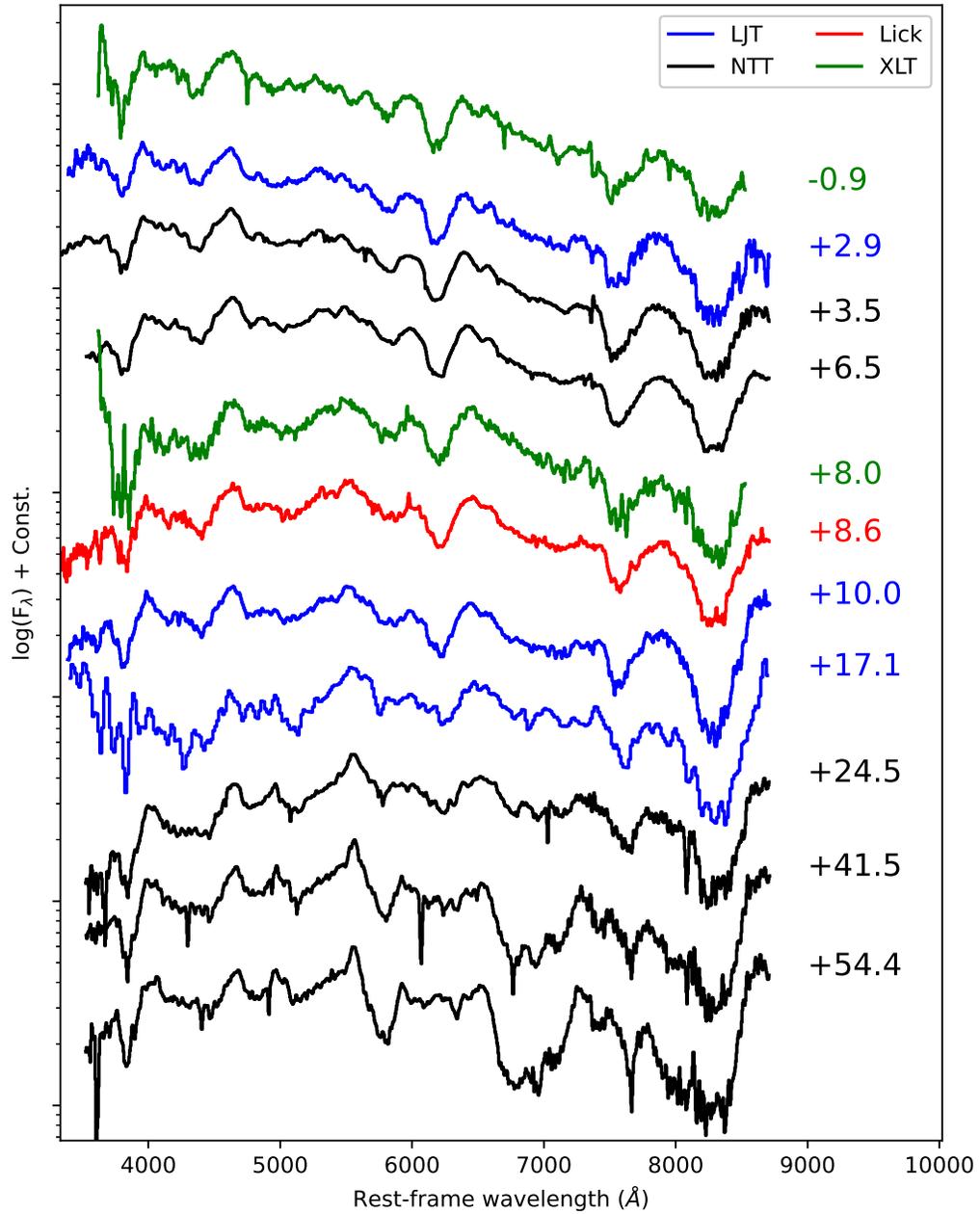}
\caption{Optical spectral evolution of SN~2016ije. The $t \approx +17$~day spectrum with a low signal-to-noise ratio is resampled with a bin size of 15~\AA, other spectra are also resampled with a bin size of 10~\AA. The spectra are shifted vertically for clarity. The epochs relative to $B$-band maximum light are labeled on the right-hand side.}
\label{fig:spectra}
\end{figure}

In Figure \ref{fig:spectra}, we display 11 optical spectra of SN~2016ije spanning phases from $t \approx -0.9$ day to $t \approx +54.4$ days. They generally share some common features with those of subluminous SNe~Ia. At $t \approx 0$ days, the spectrum was dominated by lines of IMEs, including Si~{\sc ii}, Ca~{\sc ii}, and Mg~{\sc ii}. The stronger absorptions of Si~{\sc ii} $\lambda$5972 and O~{\sc i} $\lambda$7774 with respect to normal SNe Ia indicate that the SN has a relatively low temperature and a considerable amount of oxygen was not burned, as is also favored by the presence of prominent C~{\sc ii} $\lambda$6580 absorption. In 91bg-like SNe, the presence of a significant carbon feature is infrequent, whereas in 02es-like SNe, it is a common occurrence.
The S~{\sc ii} features are not prominent, even around the maximum light.
After $t \approx 10$ days, the temperature of the continuum became progressively lower, and the absorption of Si~{\sc ii} and O~{\sc i} $\lambda$7774 weakened. After $t \approx 20$ days, absorption and emission lines of IGEs emerged in the spectra, but were weak and evolved slowly.

\begin{figure}[ht]
\centering

\includegraphics[angle=0,width=180mm]{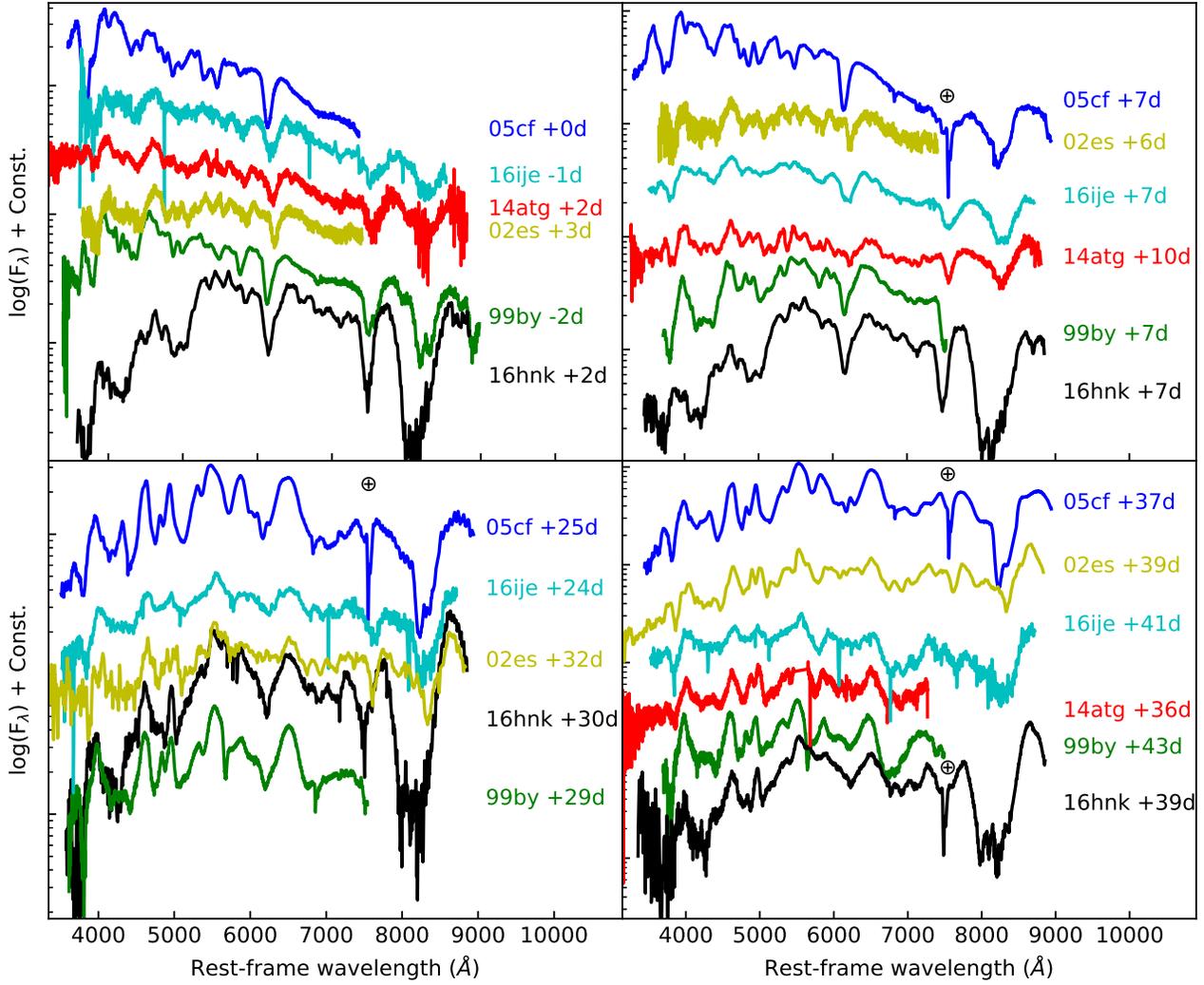}
\caption{Spectra of SN~2016ije at different epochs ($t \approx +0$, +7, +25, and +41 days) compared with spectra of subluminous SNe~Ia 1999by, 2002es, iPTF14atg, and 2016hnk, and of the normal Type Ia SN~2005cf (see the text for the references). All of the spectra have been corrected for redshift and the epochs relative to $B$-band maximum light are labeled on the right-hand side. The strongest telluric region is marked with $\oplus$.}
\label{fig:spec_compare1}
\end{figure}

To further examine the spectroscopic properties of SN~2016ije, we show detailed comparisons between SN~2016ije and other subluminous and normal SNe~Ia at several epochs ($t \approx +0$, +7, +24, and +41 days) in Figure \ref{fig:spec_compare1}. The comparison sample includes SNe 1999by \citep{gar04}, 2002es \citep{02es}, 2005cf \citep{pas07,wxf09b}, iPTF14atg \citep{cao16}, and 2016hnk \citep{gal19}, which were briefly introduced in Section \ref{sec:opticalLC}.

\subsection{Early-time Spectra}\label{sec:early_spec}

\begin{figure}[ht]
\centering
\includegraphics[angle=0,width=160mm]{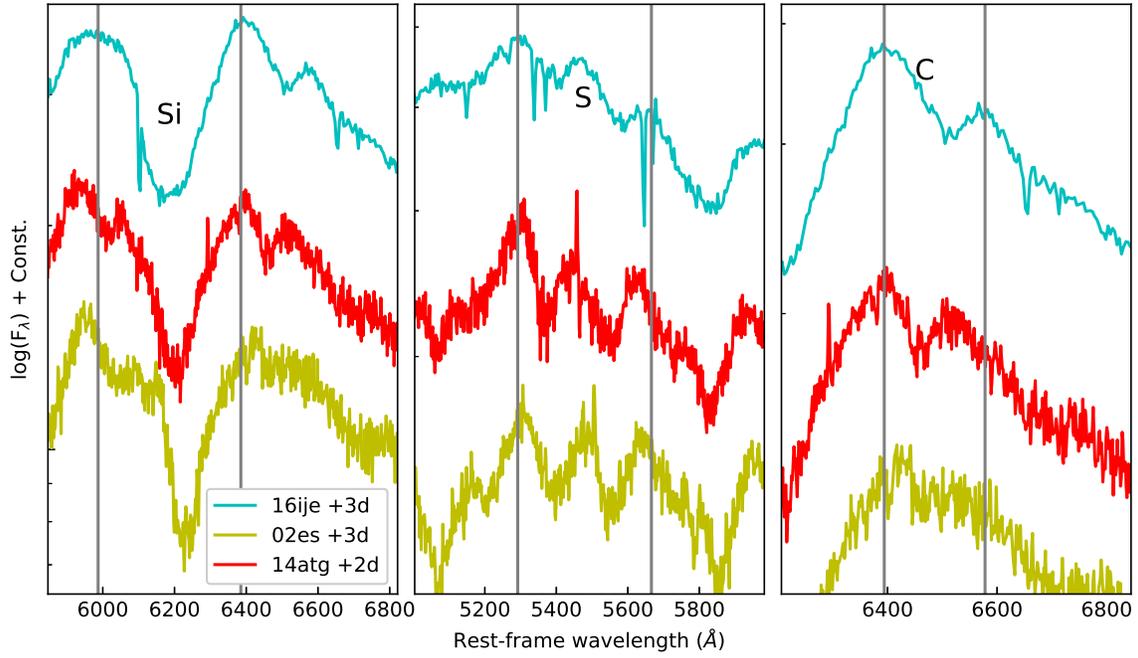}
\caption{The comparison between SNe 2002es, 2016ije and iPTF14atg in the Si~{\sc ii} $\lambda$6355, S~{\sc ii}, and C~{\sc ii} $\lambda$6580 at $t \approx +3$ days.}
\label{fig:spec_broadline}
\end{figure}

In the first week after maximum light, SN~2016ije is found to be very similar to both SN~2002es and iPTF14atg in the spectra, showing strong Si~{\sc ii} $\lambda$5972 and O~{\sc i} $\lambda$7774 lines that distinguish them from normal SNe~Ia. Compared with the two 02es-like SNe~Ia, the Ti~{\sc ii} lines are weaker but the Si~{\sc ii}, C~{\sc ii}, and S~{\sc ii} lines are much broader in SN~2016ije (as shown in Figure \ref{fig:spec_broadline}). The S~{\sc ii} lines of SN~2016ije are weaker even near maximum light, and they quickly become non-detectable, while they can still be detected in the spectra of the other two 02es-like SNe~Ia one week after maximum light. Moreover, C~{\sc ii} $\lambda$6580 appears very strong and broad in SN~2016ije, and it is visible until $t \approx +10$ days after maximum light. The IGE lines near 5000~{\AA} are also broad and shallow, which makes the absorption lines blend and become indistinguishable. All of these broader absorption lines suggest a wider line-forming region of SN 2016ije, which usually implies an extended density distribution of the ejecta \citep{ran20}.

SN~2016hnk shows somewhat similar light curves to those of SN~2016ije, but these two SNe have different shapes of continuum and absorption features (e.g., IGE lines near 5000 and 7000 {\AA}) in their early-phase spectra. For SN 2016hnk, the observed redder continuum of the spectra can be attributed to light scattering by its dense dust environment \citep{gal19}. However, the large differences that are revealed for the absorption spectral features of SN 2016hnk and SN 2016ije imply that the explosion physics of these two SNe Ia may be intrinsically different.

\subsection{Late-time Spectra}\label{sec:latespec}

\begin{figure}[ht]
\centering
\includegraphics[angle=0,width=160mm]{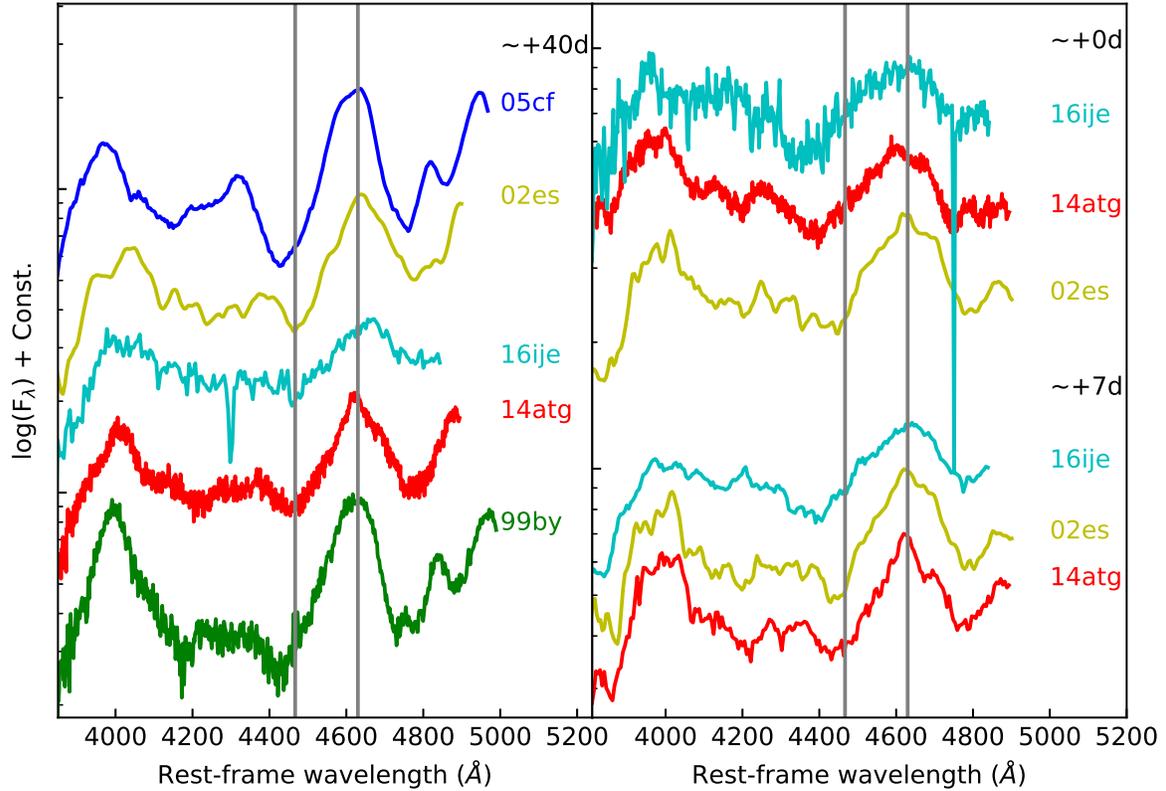}
\caption{The blended lines near 4600~{\AA} at $t \approx +40$ days (left-hand panel) and in the week of maximum light (right-hand panel). Broad absorption features are found at all phases.}
\label{fig:spec_IGE}
\end{figure}

After $ t \approx 1$ month, the spectra were dominated by IGE lines, especially Fe~{\sc ii}. SN~2016ije showed a bluer continuum and different morphology of IGE lines when compared with SN~2002es and iPTF14atg. We compared the blended iron lines near 4630~{\AA} between two gray lines in Figure \ref{fig:spec_IGE}, which look shallow and asymmetric. One can see that the morphology of the spectra near 4600~{\AA} is affected by the blue wing of iron emission lines near 4630~{\AA} and the red wing of absorption lines near 4400~{\AA}. Since we noticed that this line feature is also broader than in other 02es-like SNe~Ia at early phases (as shown in the right panel of Figure \ref{fig:spec_IGE}), we prefer to believe that it was caused by the broader red wing of the absorption lines near 4400~{\AA}.

The IGE absorption lines in the spectra of SN~2016ije seem to be weaker and broader when compared with those in other 02es-like SNe~Ia, indicating widely distributed but not abundant IGEs in the line-forming region.

\subsection{Line Properties}\label{sec:line_properties}

\begin{figure}[ht]
\centering
\includegraphics[angle=0,width=120mm]{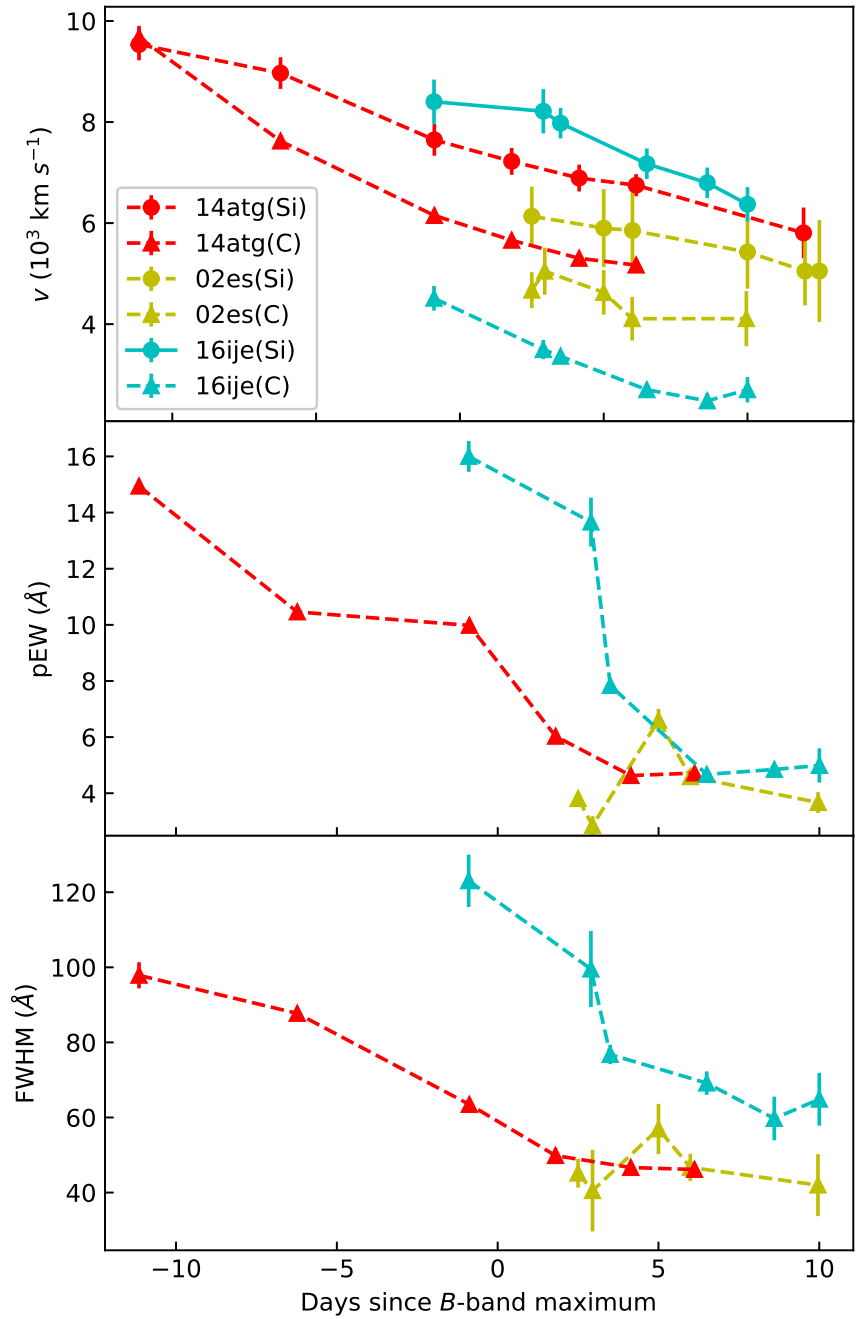}
\caption{The early-time absorption features of SNe 2002es, 2016ije, and iPTF14atg. Top: Velocities measured by the Si~{\sc ii} $\lambda$6355 and C~{\sc ii} $\lambda$6580 lines. Middle: The evolution of the pEW of C~{\sc ii} $\lambda$6580. Bottom: The evolution of the FWHM of C~{\sc ii} $\lambda$6580.}
\label{fig:vSi&carbon}
\end{figure}

Using the Si~{\sc ii} $\lambda$6355 absorption feature, we measure the photospheric velocity of SN~2016ije. It is estimated to be $v_{\rm{Si}}$ = 8300 $\pm$ 200 km s$^{-1}$ around maximum light, which is lower than that of a typical SN~Ia by about 2000 km s$^{-1}$ and even lower than that of 91bg-like SNe~Ia at a similar phase. The velocity gradient after the peak is estimated to be $-$200 $\pm$ 20 km s$^{-1}$ d$^{-1}$ for SN~2016ije, which suggests that it can be put into the high-velocity-gradient (HVG) subclass according to the classification scheme proposed by \citet{ben05}. In Figure \ref{fig:vSi&carbon}, we compare the velocity evolution of SN~2016ije with that of SN~2002es and iPTF14atg. All three 02es-like SNe~Ia exhibit relatively lower expansion velocity and rapid velocity evolution when compared with normal SNe~Ia at all phases, and $v_{\rm{Si}}$ of SN~2016ije is the highest among these 02es-like SNe.

We also measured the velocity inferred from C~{\sc ii} $\lambda$6580 absorption, with $v_{\rm{C}}$ = 4500 $\pm$ 200 km s$^{-1}$ of SN 2016ije around maximum light. Among the three 02es-like SNe~Ia, SN~2016ije has the lowest carbon velocity, with a trend contrary to the Si~{\sc ii} velocity. We notice that the 02es-like SNe~Ia with high $v_{\rm{Si}}$ like SN~2006bt and PTF10ops hardly show the C~{\sc ii} $\lambda$6580 absorption features in their near-maximum-light spectra. 

The carbon features, which reflect the amount of unburned material in the explosion of C-O white dwarfs, have been qualitatively shown to be broader in SN~2016ije in section \ref{sec:early_spec}. We show the evolution of the pseudo-equivalent width (pEW) and the full width at half maxima (FWHM) of C~{\sc ii} $\lambda$6580 in Figure \ref{fig:vSi&carbon}. In addition to SN~2016ije, our sample includes SN~2002es and iPTF14atg, which are both 02es-like SNe~Ia with strong carbon lines. One can see that the C~{\sc ii} $\lambda$6580 line is also stronger in SN~2016ije, with the width being about 1.5 times that of other 02es-like SNe, which indicates that the unburned carbon is abundant and widely distributed in the explosion of SN~2016ije.

\section{Discussion}\label{sec:discussion}

\subsection{Promising Models}

\begin{figure}[ht]
\centering
\includegraphics[angle=0,width=180mm]{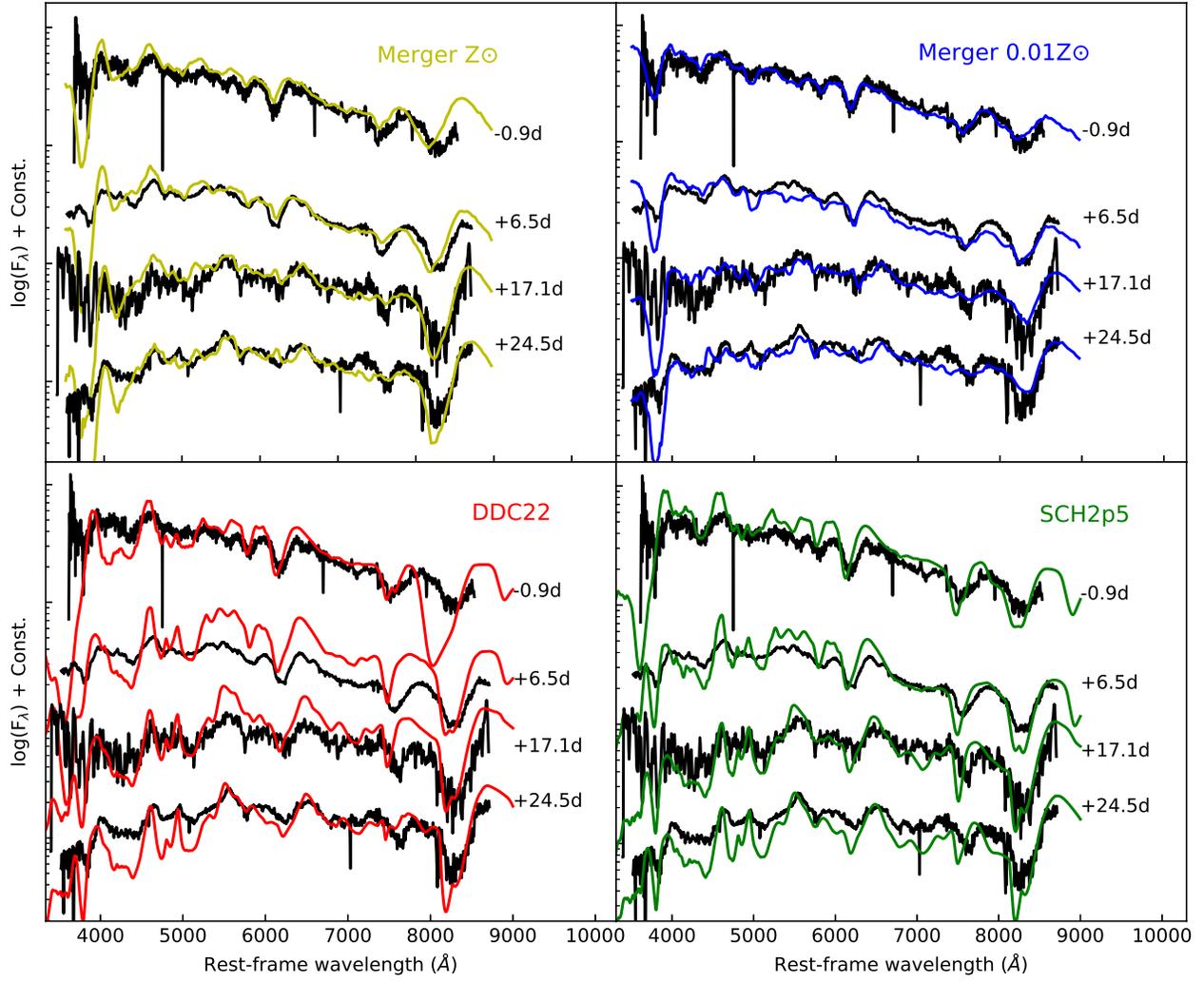}
\caption{Optical spectra of SN~2016ije compared to synthetic spectra of the violent merger model with metallicity at $Z_{\odot}$ (yellow; \citealt{kro13}) and 0.01$Z_{\odot}$ (blue; \citealt{kro16}), model SCH2p5 (red; \citealt{blond17}) and model DDC22 (green; \citealt{blond13}).}
\label{fig:blond_spec1}
\end{figure}

The violent merger model is a promising model for 02es-like SNe \citep{mag11,02es}, and some simulations also show that the synthetic spectra can fit the observations of 02es-like SN~2010lp and iPTF14atg well \citep{kro13,kro16}. We compare the spectra of SN~2016ije with the 0.9~$M_{\odot}$ and 0.76~$M_{\odot}$ violent merger models at 1~$Z_\odot$ and 0.01~$Z_\odot$ in Figure \ref{fig:blond_spec1}. Note that synthetic spectra are angle-averaged spectra, which cannot accurately describe the asymmetric properties of the violent merger. Considering that the delayed-detonation and double-detonation models can reproduce most of normal and subluminous SNe Ia, we also include the synthesis spectra of a Chandrasekhar-mass ($M_{\rm{Ch}}$) delayed-detonation model (DDC22; \citealt{blond13}) and a sub-$M_{\rm{Ch}}$ pure detonation model (SCH2p5; \citealt{blond17}) for comparison. The SCH model can be considered to be a simplification of a sub-$M_{\rm{Ch}}$ double detonation with a thin helium shell, and the reason for choosing SCH2p5 and DDC22 is that their synthetic magnitudes closely match the observed magnitudes of SN 2016ije.

Although the synthetic angle-averaged spectra from the violent merger model cannot fully reproduce the low-velocity carbon absorption line and broad absorption features, this model still fits the observations well. Compared with the violent merger model at $Z_\odot$, the metal-poor model matches the early-time spectrum of SN~2016ije better, which is consistent with the fact that SN~2016ije arose from a subsolar-metallicity stellar environment. Moreover, the double WDs in the progenitor system of the violent merger model can evolve from different paths \citep{wb18}, of which the delay-time distributions (DTDs) show that the double-degenerate system can emerge in galaxies of ages ranging from 0.1 to 10~Gyr \citep{liu16}. Wide DTDs are difficult to use as a constraint but would suggest that the double-degenerate progenitors are not in conflict with the young host galaxy of SN~2016ije. In contrast, neither the DDC nor SCH models can satisfactorily reproduce the spectral evolution of SN 2016ije, despite their ability to reproduce some normal and 91bg-like SNe Ia.

\begin{figure}[ht]
\centering
\includegraphics[angle=0,width=140mm]{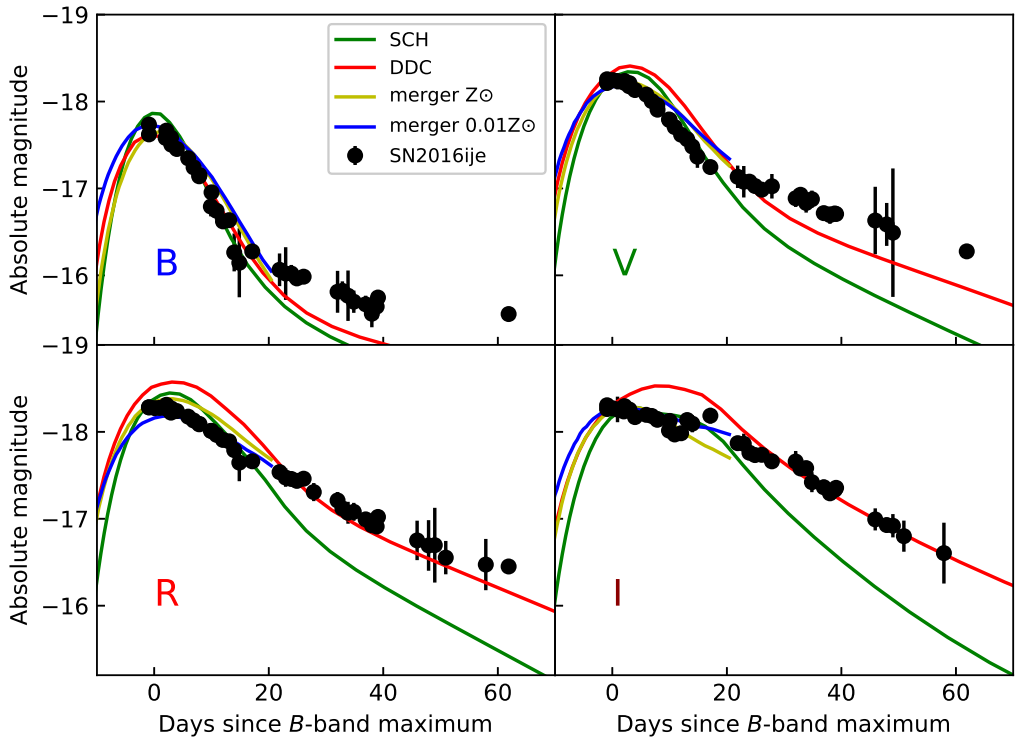}
\caption{Light curves of SN~2016ije (black circles) compared to synthetic light curves of the violent merger model with metallicity at $Z_{\odot}$ (yellow; \citealt{kro13}) and 0.01$Z_{\odot}$ (blue; \citealt{kro16}), model SCH2p5 (red; \citealt{blond17}) and model DDC22 (green; \citealt{blond13}).}
\label{fig:blond_lc}
\end{figure}

We also compare the light curve of SN~2016ije with the same models in Figure \ref{fig:blond_lc}. 
The light curves of the two violent merger models were obtained by convolution using their spectra. Additionally, the synthetic spectra of the violent merger were limited to only up to $t$ $\approx$ +20 days, which made it impossible to compare them with late-time observations.
One can see that the violent merger models and the SCH model fit the data well in the early phase, while neither SCH nor DDC model fits the late-time data satisfactorily. As \citet{kro16} well fitted the light curves of iPTF14atg with the violent merger at 0.01~$Z_\odot$, and the light curves of iPTF14atg and SN 2016ije are quite similar, the violent merger model with metallicity of 0.01~$Z_\odot$ is also very likely to fit SN 2016ije.

\subsection{Observational Peculiarities}

Our analysis indicates that SN 2016ije shares many post-maximum observed properties with 02es-like SNe Ia, particularly with iPTF14atg. However, the object still shows distinct peculiarities, especially in spectral features. As we showed in section \ref{sec:line_properties}, the FWHM and pEW of the C~{\sc ii} $\lambda$6580 absorption feature of SN~2016ije are 1.5 times those of SN~2002es and iPTF14atg. Furthermore, most other absorption lines are also broader than those of other 02es-like SNe. For SN~2016ije, although the velocity of the Si~{\sc ii} $\lambda$6355 line is slightly higher than those of SN~2002es and iPTF14atg, its velocity of C~{\sc ii} $\lambda$6580 is the lowest among these 02es-like SNe. 

The spectral peculiarities of SN 2016ije impose many requirements on the model. The broader absorption lines indicate that the line-forming region has a more extended distribution, which requires a more extended density profile or a lower temperature of the ejecta \citep{ran20}. The extremely low velocity of the strong carbon absorption line indicates that the unburned carbon remains in the inner region of the ejecta.

These requirements may be met in the violent merger model, which can provide asymmetric ejecta, leading to great diversity in the light-curve evolution, photospheric velocities, and the fraction of different elements at different viewing angles \citep{pak10,pak12}. 
During a violent merger, the remaining material from the companion star creates an IGE cavity. Within this cavity, a significant amount of unburned carbon remains in the inner region and the IMEs distribution is extended. Thus, in the simulation \citep{bul16}, the synthetic spectrum along the viewing orientation facing the IGE cavity (hereafter we call this orientation \textbf{n3}, which was defined in \citealt{bul16}) shows similar spectral properties to SN~2016ije, showing strong and low-velocity carbon absorption feature and broad absorption lines. In this model, the explosion of the primary WD produces most of the silicon, which is mainly distributed in the outer region. The difference in the spatial distribution of silicon and carbon can explain the difference in their expansion velocities, which can be inferred from the absorption lines in the spectra.

The synthetic spectrum that is generated along \textbf{n3} exhibits even broader absorption features, such as Si~{\sc ii} and S~{\sc ii}, and weaker silicon and calcium triplet lines when compared to those observed in the spectra of SN 2016ije. This inconsistency may be explained by the actual viewing angle deviating from \textbf{n3}. When the viewing angle deviates from the equatorial plane, the simulation shows a higher mass fraction of silicon in the outer region of the ejecta (as shown in Figure 2 in \citealt{pak12}), which could partly reconcile the discrepancy between the synthetic spectrum and the observations. Furthermore, additional simulations are required to determine whether the discrepancy results from differences in viewing angles or from the initial parameters, such as mass ratio. A smaller mass ratio would result in more severe tearing of the secondary, leading to differences in nucleosynthesis and density profiles, which may also explain the discrepancies between the observations of SN 2016ije and the synthetic spectra.

\section{Conclusion}\label{sec:conclusion}

In this paper, we present and analyze photometric and spectroscopic observations of SN~2016ije, which is located in a subsolar-metallicity, low-mass, and star-forming galaxy. This SN is very similar in many respects to 02es-like SNe, especially iPTF14atg, including the light curves, color curves, and spectra. We also noticed that the color curves of 02es-like SNe follow a distinct evolution pattern, which gradually becomes red after the maximum light, and maintains almost constant color between $t \approx +20$ and $+40$ days. This unique evolution pattern can aid in distinguishing 02es-like SNe from other SNe Ia.

Although SN 2016ije has many similarities to other 02es-like SNe, we have identified peculiarities in its spectra, including broader absorption features and a lower velocity of carbon. These observations suggest the presence of significant unburned carbon in the inner region, as well as an extended line-forming region along the line of sight. These peculiarities impose many requirements on the explosion model, which can be met in a violent merger model at a particular viewing angle near the orientation along the IGE cavity (i.e., \textbf{n3}). Additionally, the simulation results obtained around this viewing angle can approximately replicate the spectral peculiarities observed in SN 2016ije.

After analyzing and comparing the observed spectra of SN 2016ije with those generated by different models, we conclude that the violent merger scenario with a metal-poor progenitor is the most likely explanation for the peculiarities observed in SN 2016ije. Moreover, we found that the DDC22 and SCH2p5 models were incapable of accurately reproducing the spectral evolution of SN~2016ije.

Although the sample size of 02es-like SNe Ia is currently small and the asymmetry of the violent merger model makes simulations complex, the upcoming Rubin Observatory's Legacy Survey of Space and Time (LSST; \citealt{LSST}) and SiTian \citep{sitian} will greatly enrich the sample of 02es-like SNe Ia, enabling a better understanding of their diversity. In the future, more accurate three-dimensional numerical simulations and polarization studies of 02es-like SNe will help determine whether the violent merger model is a suitable explosion model.

\begin{acknowledgments}

We acknowledge the support of the staff of the various observatories at which the data were obtained. This work is based (in part) on observations collected at the European Organisation for Astronomical Research in the Southern Hemisphere, Chile as part of PESSTO (the Public ESO Spectroscopic Survey for Transient Objects) ESO programs 188.D-3003 and 191.D-0935.
Funding for the LJT has been provided by Chinese Academy of Sciences and the People’s Government of Yunnan Province. The LJT is jointly operated and administrated by Yunnan Observatories and Center for Astronomical Mega-Science, CAS.
Some of the data presented herein were obtained at the W. M. Keck Observatory, which is operated as a scientific partnership among the California Institute of Technology, the University of California, and NASA; the observatory was made possible by the generous financial support of the W. M. Keck Foundation.
The Kast spectrograph on the Shane 3~m telescope at Lick Observatory was made possible through a gift from William and Marina Kast.
Research at Lick Observatory is partially supported by a generous gift from Google.

This work is partially supported by the Open Project Program of the Key Laboratory of Optical Astronomy, National Astronomical Observatories, Chinese Academy of Sciences.
This work is supported by the National Natural Science Foundation of China (NSFC; grants 12233008, 12120101003, 12288102, 12033003, and 11633002), the Scholar Program of Beijing Academy of Science and Technology (DZ:BS202002), Beijing Municipal Natural Science Foundation under grant 1222028, and the Tencent Xplorer Prize. This work is also partially supported by NSFC (grants 11890691, 11890693, 11733007, 11673027, 11873053, 11178003, 11325313, and 12073035), and China Manned Spaced Project (CMS-CSST-2021-A12). J.-J. Zhang is supported by the NSFC (grants 11403096 and 11773067), the Key Research Program of the CAS (grant KJZD-EW-M06), the Youth Innovation Promotion Association of the CAS (grant 2018081), and the CAS ``Light of West China'' Program.
B.W. is supported by the NSFC (grant 12225304), the National Key R\&D Program of China (grant 2021YFA1600404), and the Western Light Project of CAS (grant XBZG-ZDSYS-202117). L.G. acknowledges financial support from the Spanish Ministerio de Ciencia e Innovaci\'on (MCIN), the Agencia Estatal de Investigaci\'on (AEI) 10.13039/501100011033, and the European Social Fund (ESF) ``Investing in Your Future'' under the 2019 Ram\'on y Cajal program RYC2019-027683-I and the PID2020-115253GA-I00 HOSTFLOWS project, from Centro Superior de Investigaciones Cient\'ificas (CSIC) under the PIE project 20215AT016, and the program Unidad de Excelencia Mar\'ia de Maeztu CEX2020-001058-M. M.G. is supported by the EU Horizon 2020 research and innovation programme under grant agreement 101004719. A.V.F.'s supernova group at U.C. Berkeley is supported by the Christopher R. Redlich Fund and many individual donors.

We thank S. Blondin, P. A. Mazzali, C. Wu, D. Liu, W. Li, and Y. Zhang for their helpful discussions of this paper.

\software{SWARP\ \ \citep{SWARP},\ \ SExtractor\ \ \citep{SExtractor},\ \ SCAMP\ \ \citep{SCAMP},\ \ Matplotlib\ \ \citep{matplotlib},\ \ NumPy\ \ \citep{numpy},\ \ SciPy\ \ \citep{scipy}.}

\end{acknowledgments}

\clearpage

\bibliographystyle{aasjournal}
\bibliography{lizt}

\clearpage
\begin{deluxetable}{lcrcccccr}
\tabletypesize{\footnotesize}
\centerwidetable
\tablecaption{Optical Photometric Observations of SN~2016ije \label{table:photometry}}
\tablehead{\colhead{UT Date} & \colhead{MJD} & 
\colhead{Phase\tablenotemark{a}} & \colhead{$U$} & \colhead{$B$} & 
\colhead{$V$} & \colhead{$R$} & 
\colhead{$I$} & \colhead{Telescope}}
\startdata
2016 Nov. 23 &57715.679 &-0.9 &\nodata     &18.25(0.03) &17.61(0.03) &17.52(0.02) &17.50(0.05) &TNT\\
2016 Nov. 23 &57715.680 &-0.9 &17.84(0.03) &18.14(0.04)	&17.57(0.04) &17.52(0.05) &17.46(0.03) &LJT\\
2016 Nov. 24 &57716.689 &0.1  &\nodata     &\nodata 	&17.58(0.02) &17.53(0.02) &17.50(0.04) &TNT\\
2016 Nov. 25 &57717.505 &0.9  &\nodata     &\nodata 	&17.59(0.09) &17.53(0.07) &17.51(0.15) &TNT\\
2016 Nov. 26 &57718.595 &2.0  &\nodata     &18.29(0.03) &17.61(0.02) &17.53(0.02) &17.53(0.04) &TNT\\
2016 Nov. 26 &57718.724 &2.1  &17.91(0.02) &18.21(0.03)	&17.59(0.03) &17.49(0.07) &17.47(0.04) &LJT\\
2016 Nov. 27 &57719.531 &2.9  &18.01(0.02) &18.29(0.07)	&17.65(0.04) &17.58(0.03) &17.51(0.03) &LJT\\
2016 Nov. 27 &57719.586 &3.0  &\nodata     &18.37(0.02) &17.62(0.02) &17.53(0.02) &17.51(0.03) &TNT\\
2016 Nov. 28 &57720.555 &4.0  &\nodata     &18.42(0.02) &17.69(0.02) &17.56(0.02) &17.60(0.04) &TNT\\
2016 Nov. 30 &57722.528 &5.9  &\nodata     &18.53(0.03) &17.75(0.02) &17.63(0.02) &17.57(0.05) &TNT\\
2016 Dec. 1  &57723.479 &6.9  &\nodata     &18.63(0.03) &17.82(0.03) &17.67(0.03) &17.58(0.04) &TNT\\
2016 Dec. 2  &57724.451 &7.9  &\nodata     &18.74(0.02) &17.87(0.02) &17.71(0.02) &17.63(0.04) &TNT\\
2016 Dec. 2  &57724.497 &7.9  &\nodata 	   &18.72(0.06)	&17.92(0.05) &17.71(0.03) &17.62(0.03) &LJT\\
2016 Dec. 4  &57726.492 &9.9  &\nodata     &19.13(0.11) &18.03(0.05) &17.79(0.05) &17.75(0.08) &TNT\\
2016 Dec. 4  &57726.610 &10.0 &18.96(0.02) &18.92(0.06)	&18.04(0.05) &17.79(0.03) &17.64(0.03) &LJT\\
2016 Dec. 5  &57727.492 &10.9 &\nodata     &19.13(0.10) &18.12(0.05) &17.84(0.04) &17.80(0.04) &TNT\\
2016 Dec. 6  &57728.610 &12.0 &\nodata     &19.25(0.05) &18.21(0.03) &17.89(0.02) &17.78(0.04) &TNT\\
2016 Dec. 7  &57729.671 &13.1 &19.51(0.03) &19.24(0.05) &18.26(0.03) &17.91(0.04) &17.63(0.04) &LJT\\
2016 Dec. 8  &57730.567 &14.0 &\nodata     &19.61(0.22) &18.34(0.09) &18.01(0.07) &17.67(0.10) &TNT\\
2016 Dec. 9  &57731.460 &14.9 &\nodata     &19.73(0.40) &18.46(0.13) &18.16(0.21) &\nodata     &TNT\\
2016 Dec. 11 &57733.719 &17.1 &\nodata     &19.60(0.08)	&18.58(0.03) &18.14(0.04) &17.58(0.09) &LJT\\
2016 Dec. 16 &57738.487 &21.9 &\nodata     &19.81(0.19) &18.69(0.13) &18.26(0.08) &17.90(0.06) &TNT\\
2016 Dec. 17 &57739.555 &23.0 &\nodata     &19.85(0.30) &18.75(0.18) &18.33(0.11) &17.90(0.12) &TNT\\
2016 Dec. 18 &57740.540 &23.9 &\nodata     &19.85(0.10) &18.75(0.06) &18.34(0.04) &18.01(0.06) &TNT\\
2016 Dec. 19 &57741.511 &24.9 &\nodata     &19.91(0.08) &18.80(0.05) &18.36(0.04) &18.03(0.06) &TNT\\
2016 Dec. 20 &57742.692 &26.1 &\nodata     &19.89(0.04)	&18.84(0.04) &18.34(0.03) &18.03(0.03) &LJT\\
2016 Dec. 22 &57744.464 &27.9 &\nodata     &\nodata 	&18.80(0.14) &18.49(0.10) &18.11(0.07) &TNT\\
2016 Dec. 26 &57748.617 &32.0 &\nodata     &20.06(0.24) &18.94(0.10) &18.59(0.10) &18.11(0.12) &TNT\\
2016 Dec. 27 &57749.459 &32.9 &\nodata     &20.05(0.11) &18.90(0.07) &18.68(0.06) &18.19(0.07) &TNT\\
2016 Dec. 28 &57750.445 &33.8 &\nodata     &20.11(0.29) &18.99(0.11) &18.73(0.13) &18.18(0.07) &TNT\\
2016 Dec. 29 &57751.450 &34.8 &\nodata     &20.18(0.13) &18.95(0.10) &18.72(0.10) &18.34(0.12) &TNT\\
2016 Dec. 31 &57753.523 &36.9 &\nodata     &20.21(0.10) &19.11(0.08) &18.81(0.06) &18.40(0.09) &TNT\\
2017 Jan. 1  &57754.605 &38.0 &\nodata     &20.32(0.16) &19.13(0.10) &18.88(0.08) &18.47(0.08) &TNT\\
2017 Jan. 2  &57755.445 &38.8 &\nodata     &20.23(0.14) &19.11(0.08) &18.89(0.06) &18.43(0.05) &TNT\\
2017 Jan. 2  &57755.674 &39.1 &\nodata     &20.13(0.03)	&19.12(0.05) &18.78(0.03) &18.41(0.03) &LJT\\
2017 Jan. 8  &57762.510 &45.9 &\nodata     &\nodata     &19.20(0.39) &19.05(0.23) &18.77(0.13) &TNT\\
2017 Jan. 10 &57764.511 &47.9 &\nodata     &\nodata     &19.24(0.25) &19.11(0.29) &18.84(0.08) &TNT\\
2017 Jan. 11 &57765.594 &48.9 &\nodata     &\nodata     &19.34(0.74) &19.10(0.43) &18.85(0.13) &TNT\\
2017 Jan. 13 &57767.511 &50.9 &\nodata     &\nodata     &\nodata     &19.25(0.19) &18.97(0.18) &TNT\\
2017 Jan. 17 &57771.075 &54.5 &\nodata     &20.364(0.026) &19.541(0.017) &19.237(0.019) &18.937(0.022) &NTT\\
2017 Jan. 20 &57774.504 &57.9 &\nodata     &\nodata     &\nodata     &19.33(0.30) &19.16(0.35) &TNT\\
2017 Jan. 24 &57778.491 &61.9 &\nodata     &20.32(0.05)	&19.55(0.06) &19.35(0.04)	&\nodata   &LJT\\
\enddata
\tablenotetext{a}{Relative to the epoch of $B$-band maximum (MJD = 57716.6) in the observer's frame.} 
\end{deluxetable}

\clearpage

\clearpage

\begin{deluxetable}{lcrcccr}
\tabletypesize{\footnotesize}
\centerwidetable
\tablecaption{NIR Photometric Observations of SN~2016ije \label{table:photometry_nir}}
\tablehead{\colhead{UT Date} & \colhead{MJD} & 
\colhead{Phase\tablenotemark{a}} & \colhead{$J$} & \colhead{$H$} & 
\colhead{$K$} & \colhead{Telescope}}
\startdata
2016 Nov. 27 &57719.043 &2.4  &17.465(0.345) &17.251(0.235) &17.239(0.410) &NTT\\
2016 Dec. 2  &57724.070 &7.5  &17.603(0.328) &17.240(0.228) &17.223(0.208) &NTT\\
2016 Dec. 20 &57742.083 &25.5 &17.722(0.398) &17.469(0.259) &17.995(0.232) &NTT\\
2016 Dec. 27 &57749.030 &32.4 &18.601(0.286) &\nodata &\nodata &NTT\\
2016 Jan. 4  &57757.113 &40.5 &18.615(0.373) &18.063(0.248) &19.247(0.441) &NTT\\
2016 Jan. 18 &57772.070 &55.5 &19.609(0.259) &19.225(0.344) &19.174(0.468) &NTT\\
\enddata
\tablenotetext{a}{Relative to the epoch of $B$-band maximum (MJD = 57716.6) in the observer's frame.} 
\end{deluxetable}

\begin{deluxetable}{lcrlcl}
\tabletypesize{\small}
\tablecaption{\label{table:spec}Journal of Spectroscopic Observations of SN~2016ije}
\tablehead{\colhead{UT Date} & \colhead{MJD} & 
\colhead{Phase\tablenotemark{a}} & \colhead{Range(\AA)} & 
\colhead{Resolution(\AA)\tablenotemark{b}} & \colhead{Instrument}}
\startdata
2016 Nov. 24& 57715.7& -0.9& 3700--8800& 2.8& XLT BFOSC\\
2016 Nov. 27& 57719.1& 2.5& 9400--16,400& 6.9& NTT SOFI\\
2016 Nov. 27& 57719.5& 2.9& 3500--9100& 2.9& LJT YFOSC\\
2016 Nov. 28& 57720.1& 3.5& 3300--10,000& 4.1& NTT EFOSC\\
2016 Dec. 1 & 57723.1& 6.5& 3600--9200& 5.5& NTT EFOSC\\
2016 Dec. 2 & 57724.6& 8.0& 3700--8800& 2.8& XLT BFOSC\\
2016 Dec. 4 & 57725.2& 8.6& 3400--10,400& 2.0& Lick Kast\\
2016 Dec. 5 & 57726.6& 10.0& 3500--9100& 2.9& LJT YFOSC\\
2016 Dec. 11& 57733.7& 17.1& 3500--9100& 2.9& LJT YFOSC\\
2016 Dec. 19& 57741.1& 24.5& 3600--9200& 5.5& NTT EFOSC\\
2017 Jan. 5 & 57758.1& 41.5& 3600--9200& 5.5& NTT EFOSC\\
2017 Jan. 18& 57771.0& 54.4& 3600--9200& 5.5& NTT EFOSC\\
2017 Oct. 14& 58040.6& \nodata\tablenotemark{c}& 4410--9630& 0.65& Keck II+DEIMOS\\
\enddata
\tablenotetext{a}{Relative to the epoch of $B$-band maximum (MJD = 57716.6) in the frame of the observer.}
\tablenotetext{b}{Approximate spectral resolution (FWHM).}
\tablenotetext{c}{The spectrum of the host galaxy.}
\end{deluxetable}

\begin{deluxetable}{ll}
\tabletypesize{\small}
\tablecaption{\label{table:obj}Parameters of SN~2016ije}
\tablehead{\colhead{Parameter} & \colhead{Value}}
\startdata
R.A.(J2000)     & 01$^{\rm{h}}$58$^{\rm{m}}$30$^{\rm{s}}$.33 \\
Decl.(J2000)    & +12$^{\circ}$55$\arcmin$27$\arcsec$9 \\
Date of $B_{\rm{max}}$(MJD)      & 57716.6 $\pm$ 0.8   \\
$E(B-V)_{\rm{MW}}$   & 0.0462 mag  \\ 
${\Delta}m_{15}(B)$  & 1.35 $\pm$ 0.14 mag  \\
Redshift             & 0.0324 $\pm$ 0.0003\\
Distance modulus     & 35.68 $\pm$ 0.04 mag      \\ 
$m_{B,\rm{max}}$     & 18.22 $\pm$ 0.04 mag     \\ 
$M_{B,\rm{max}}$     &  $-$17.65 $\pm$ 0.06 mag     \\ 
$s_{\rm{BV}}$            &   0.57$\pm$ 0.06 \\
\enddata
\end{deluxetable}

\clearpage

\end{document}